\documentstyle[11pt]{article}
\newcommand{\be}{\begin{equation}}
\newcommand{\ee}{\end{equation}}

\title{REMARKS ON WHITHAM AND RG}
\author{Robert Carroll\\University of Illinois\\
email:  rcarroll@math.uiuc.edu}

\date{January, 1998}

\begin{document}
\bibliographystyle{plain}
\maketitle
\begin{abstract}
We collect a number of facts and conjectures concerning Whitham theory
and the renormalization group (RG).  Some explicit relations and problems
are indicated in
the context of $N=2$ susy Yang-Mills (YM).
\end{abstract}

\section{INTRODUCTION}
\renewcommand{\theequation}{1.\arabic{equation}}\setcounter{equation}{0}

Relations between Whitham equations and RG in Seiberg-Witten (SW) theory
have been suggested in many places (see e.g. 
\cite{ea,gc,gd,ge,gg,ia,md,mf,na,tc}) without an explicit unification
or clarification of roles.
We do not claim to achieve the latter but we will indicate what seem to
be some paths in this direction along with some problems. 
Let us begin with the original
$SU(2)$ Seiberg-Witten (SW) curves (cf. \cite{sd}).  Thus for
$N=2$ SYM (supersymmetric Yang-Mills theory) the moduli space of
quantum vacua is the $u$ plane with singularities at $-1,1,\infty$ and
a ${\bf Z_2}$ symmetry $u\to -u$  (we will replace the $\pm 1$ with
a scaling factor $\pm \Lambda^2$ later).  Over the punctured $u$ plane
there is a flat $SL(2,{\bf R})$ bundle $V$ with the following
monodromies around $\infty,1,-1$:
\be
M_{\infty}=\left(\begin{array}{cc}
-1 & 2\\
0 & -1
\end{array}\right);\,\,M_1=\left(\begin{array}{cc}
1 & 0\\
-2 & 1
\end{array}\right);\,\,M_{-1}=\left(\begin{array}{cc}
-1 & 2\\
-2 & 3
\end{array}\right)
\label{A}
\ee
The quantities $(a^D(u),a(u))^T$ are a holomorphic section of $V\otimes
{\bf C}$ with asymptotic behavior $a\sim \sqrt{2u},\,\,a^D\sim i(\sqrt{2u}/\pi)
log(u)$ near $\infty$ and $a^D\sim c_0(u-1),\,\,a\sim a_0+(i/\pi)a^Dlog(a^D)$
near $u=1$ (near $u=-1$ the behavior is similar with $a-a^D$ replacing
$a^D$).  The monodromy matrices generate a subgroup $\Gamma(2)\subset
SL(2,{\bf R})$ and one can represent the moduli space as 
${\cal M}=H/\Gamma(2)$
where $H$ is the Poincar\'e upper half plane.  The family of curves
parametrized by ${\cal M}$ (SW curves) is given by
\be
y^2=(x-1)(x+1)(x-u)
\label{B}
\ee
so that over each $u\in {\cal M}$ there is a genus one Riemann surface
(RS) $E_u$ determined by (\ref{B}).  One defines differentials
$\lambda_1=dx/y$ (holomorphic)
and $\lambda_2=xdx/y$ and chooses a suitable
basis of one cycles $(\gamma_1,\gamma_2)\sim (A,B)$ (e.g. take $\gamma_1\sim
-1\to 1$ followed by $1\to -1$ and $\gamma_2\sim 1\to u$ followed
by $u\to 1$).  Then the SW differential is defined via
$\lambda_{SW}=(1/\pi\sqrt{2})(\lambda_2-u\lambda_1)$ and the quantities
$a,a^D$ are taken as 
\be
a=\oint_A\lambda_{SW}=\frac{\sqrt{2}}{\pi}\int_{-1}^1
\frac{dx\sqrt{x-u}}{\sqrt{x^2-1}};\,\,a^D=\oint_B\lambda_{SW}=
\frac{\sqrt{2}}{\pi}\int_1^u\frac{dx\sqrt{x-u}}{\sqrt{x^2-1}}
\label{C}
\ee
Next one remarks that the $\tau$ parameter of the elliptic curve $E_u$
can be written as $\tau=(da^D/du/da/du)\sim da^D/da$ and hence
$\Im\tau>0$.
\\[3mm]\indent
Now this curve was examined in \cite{gc} in connection with elliptic
one gap solutions of KdV \`a la Gurevich-Pitaevskij (GP) \cite{gk}.  This is 
partially summarized as follows.  First one computes
\be
a=\oint_A\lambda_{SW}=-2i\psi_1(u)=\sqrt{2}(u+1)^{1/2}F\left(-\frac{1}{2},\frac
{1}{2},1,\frac{2}{u+1}\right);
\label{ZL}
\ee
$$a^D=\oint_B\lambda_{SW} =i\psi_2=i\frac{u-1}{2}F
\left(\frac{1}{2},\frac{1}{2},
2,\frac{1-u}{2}\right)$$
where $(-\partial_z^2+W(z))\psi=0$ with $W=-(1/4)[1/(z^2-1)] \,\,(z\sim x)$.
The GP solution of \cite{gk} is an elliptic
one gap solution to KdV, namely (${\cal P}\sim$ Weierstrass function)
\be
\tilde{u}(t_1,t_3,\cdots|u)=\frac{\partial^2}{\partial t_1^2}log\tau
(t_1,t_3,\cdots|u)=u_0{\cal P}(k_1t_1+k_3t_3+\cdots+\Phi_o|\omega,\omega')
+\frac{u}{3}
\label{FK}
\ee
Here one recalls that for KdV there are differentials 
$\Omega_n,\,\,n>0$
\be
d\Omega_{2j+1}(z)=\frac{P_{j+g}(z)}{y(z)}dz;\,\,y^2\sim (z^2-1)(z-u)
\label{FL}
\ee
In particular one writes
\be
dp\equiv d\Omega_1=\frac{z-\alpha(u)}{y(z)}dz;\,\,dE\equiv d\Omega_3(z)
=\frac{z^2-\frac{1}{2}uz-\beta(u)}{y(z)}dz
\label{FM}
\ee
The normalization conditions $\oint_A d\Omega_i=0$ yield
$\alpha(u)$ and $\beta(u)$ immediately.  Associated with this situation
we have the classical Whitham theory (cf.
\cite{Aa,bb,cc,cg,ci,cj,fc,fb,gb,ia,kb,kc,ko,na,ta,tc,td}) giving ($t_n\to T_n=
\epsilon t_n,\,\,\epsilon\to 0$)
\be
\frac{\partial d\Omega_i(z)}{\partial T_j}=\frac{\partial d\Omega_j(z)}
{\partial T_i};\,\,d\Omega_i(z)=\frac{\partial dS(z)}{\partial T_i}
\label{FN}
\ee
where $dS$ is some action term which classically was thought of
in the form $dS=\sum T_id\Omega_i$.  In fact it will continue to have
such a form in any context for generalized times $T_A$ (including the
$a_i$) and generalized differentials $d\Omega_A$ (cf. \cite{ka,ko}).  
Further taking coordinates
$u_{\alpha}$ as the branch points of the corresponding hyperelliptic
(here elliptic) surface one has the hydrodynamic type equations
\be
\frac{\partial u_{\alpha}}{\partial T_i}=v_{ij}^{\alpha\beta}(u)\frac
{\partial u_{\beta}}{\partial T_j};\,\,v_{ij}^{\alpha\beta}=\delta^
{\alpha\beta}\left.\frac{d\Omega_i(z)}{d\Omega_j(z)}\right|_{z=u_{\alpha}}
\label{FO}
\ee
\indent
Now what happens is that after one switches on the Whitham dynamics
the periods of the differential $d{\cal S}$ become the periods of
the ``modulated" function in (\ref{FK}).  To be more precise it is shown
in \cite{gc} that 
\be
dS(z)=\left(T_1+T_3(z+\frac{1}{2}u)+\cdots\right)\times
\frac{z-u}{y(z)}dz=g(z|T_i,u)\lambda_{SW}(z)
\label{FPP}
\ee
where $\lambda_{SW}$ is the SW differential $(z-u)dz/y(z)$.
The demonstration is sort of ad hoc and goes as follows.  
Setting $T_{2k+1}=0$ for $k>1$ and computing from
(\ref{FPP}) one gets
\be
\frac{\partial dS(z)}{\partial T_1}=\left(z-u-(\frac{1}{2}T_1+\frac{3}{4}uT_3)
\frac{\partial u}{\partial T_1}\right)\frac{dz}{y(z)};
\label{FQQ}
\ee
$$\frac{\partial dS(z)}{\partial T_3}=\left(z^2-\frac{1}{2}uz-\frac{1}{2}u^2
-(\frac{1}{2}T_1+\frac{3}{4}uT_3)\frac{\partial u}
{\partial T_3}\right)\frac{dz}{y(z)}$$
and comparing with (\ref{FM}) gives
\be
(\frac{1}{2}T_1+\frac{3}{4}uT_3)\frac{\partial u}
{\partial T_1}=\alpha(u)-u;\,\,(\frac
{1}{2}T_1+\frac{3}{4}uT_3)\frac{\partial u}
{\partial T_3}=\beta(u)-\frac{1}{2}u^2
\label{FR}
\ee
Hence the construction gives a solution to the general Whitham equation
of the form
$\partial u/\partial T_3=v_{31}(u)(\partial u/\partial T_1)$ with
\be
v_{31}=\frac{\beta(u)-\frac{1}{2}u^2}{\alpha(u)-u}=\left.\frac
{d\Omega_3(z)}{d\Omega_1(z)}\right|_{z=u}
\label{FFS}
\ee
which is what it should be from the general Whitham theory (cf. (\ref{FO})).
It follows that
\be
a=\left.\frac{1}{T_1}\oint_AdS(z)\right|_{T_3=T_5=\cdots=0};\,\,
a_D=\left.\frac{1}{T_1}\oint_BdS(z)\right|_{T_3=T_5=\cdots=0}
\label{FT}
\ee
and also
\be
\frac{\partial}{\partial T_i}\oint_AdS=\oint_Ad\Omega_i=0;\,\,
\frac{\partial}{\partial T_i}\oint_BdS=\oint_Bd\Omega_i=k_i
\label{FU}
\ee
where the $k_i$ are the frequencies in the original KdV solution (\ref{FK}).
We note that in (\ref{FT}) $\left.(1/T_1)dS\right|_{T_3=0}=[(z-u)/y(z)]dz=
\lambda_{SW}$ is fine but one does not have the form 
$d{\cal S}=ad\omega+\sum T_nd\Omega_n$ as in \cite{ia}
(cf. also \cite{cc,cg}) where
$cd\omega=dz/y(z)=dv$ is the canonical holomorphic differential with
$\oint_Ad\omega=1$ and $\oint_Bd\omega=\tau$ (note $c=c(u)$).  
It is at this point that
one appreciates the subtlety of the argument in \cite{ia}
expressing $d{\cal S}$ as $ad\omega+\sum T_nd\Omega_n$
but \cite{gc} provides the invaluable service of exhibiting connections
to Whitham and showing different roles for Whitham times (cf. also \cite{ea}).
\\[3mm]\indent
{\bf REMARK 1.1.}$\,\,$  The formula (\ref{FT}) suggests a normalization
$T_1\sim 1$, or better, with scaling factors $\Lambda^2$ inserted as in
\cite{ea,ia}, $T_1\sim (\sqrt{2}/\pi)\Lambda$ (cf. Remark 1.2 and
Section 6).
\\[3mm]\indent
In order to introduce a prepotential one can compare here to \cite{ia}
where $d{\cal S}_{min}\sim\lambda_{SW}$ with 
$\partial d{\cal S}_{min}/\partial u=-(1/2\pi\sqrt{2})(dz/y)=
-(1/2\pi\sqrt{2})c(u)d\omega$.  Note also that $\oint_A\sim2\int_{-1}^1$
and $\oint_B\sim 2\int_1^u$.  Then $F_{red}(a)$ is defined
as $F(a,T_n)$ for $T_n=0$ when $n>1$ or $n<-1$.  Note here
that a Toda theory with times $T_0,\,\,T_{\pm n}\,\,(n\geq 1)$ is used
in \cite{ia} with two points $P_{\pm}\sim\infty$
to represent the SW elliptic curve
(this is sketched below); the approach of
\cite{gc} sketched above uses a KP 
(or KdV) format with $T_n\,\,(N\geq 1)$ and we
saw that $T_n=0$ for $n>1$ with $T_1=1$ could be used in describing
$\lambda_{SW}$.
In \cite{ia} this leads to $F_{red}=(1/2)aa^D-(iu/\pi)$ so
$F_{red}$ is not homogeneous whereas $(\bullet)\,\,2F=
\sum T_nd\Omega_n+\sum a_id\omega_i=\sum T_n\partial_nF+\sum a_i\partial
F/\partial a_i$ is homogeneous of degree two (cf. also \cite{cc}).  
This is also 
developed in \cite{ea} for example in a semi-Toda format
where for susy YM coupled to massless
hypermultiplets ($\sim T_0=0$) a basic formula is
\be
aa^D-2F_{red}=-T_1\partial_1F_{red}=8\pi ib_1u
\label{ZM}
\ee
Thus $F_{red}=(1/2)aa^D-4\pi ib_1u\Rightarrow (i/\pi)=4\pi ib_1$ or 
$b_1=(1/4\pi^2)$ (cf. also \cite{ma}).  Note in (\ref{FT}) etc. in a KP
format we could think of $F_{red}$ in the same way since for 
$T_1=1$ and $T_n=0\,\,(n>1)$ (\ref{FPP}) is $dS=\lambda_{SW}=
d{\cal S}_{min}$.  Now the definition of $a^D$ as $\partial F_{red}/
\partial a$ implies
\be
a^D=\frac{1}{2}a^D+\frac{1}{2}a\frac{\partial a^D}{\partial a}+
\frac{1}{i\pi}\frac{\partial u}{\partial a};\,\,
\frac{\partial a}{\partial u}=-\frac{1}{\pi\sqrt{2}}\int_{-1}^1
\frac{dz}{y}=-\frac{c(u)}{2\pi\sqrt{2}}
\label{ZN}
\ee
Then e.g. $(1/2)a^D=(1/2)a(\partial a^D/\partial a)+(1/i\pi)
(\partial u/\partial a)$ and $\partial_aa^D=\partial_a\oint_B
d{\cal S}_{min}=\oint_Bd\omega=\tau(u)=(4\pi i/g^2)+(\theta/2\pi)$
where the important objects
$g,\,\,\theta$ are functions of $a$ or $u$.  
\\[3mm]\indent
Now the $b_1$ term in (\ref{ZM}) is related to renormalization (see
e.g. \cite{bc,By,ba,Bz,be,ea,ff,gc,he,Ka,ma,na,se}).  In \cite{na} for
example it was shown that the SW solution corresponds to Whitham dynamics
when the prepotential $F$ satisfies the homogeneity condition
$(\bullet)$ in the form $aF_a-2F+\sum T_n\partial_nF=0$.  For situations
with massless matter fields where $T_0=0$ the procedure of \cite{ea}
involves putting $T_n=0$ except for $T_1$ and showing that $-T_1\partial_1F
=8\pi ib_1u$ as in (\ref{ZM}).  Here a general curve 
$y^2=(x^2-\Lambda^4)(x-u)$ is used ($\Lambda$ being a scaling factor)
and $b_1$ is the coefficient of the 1-loop beta function.  
(cf. \cite{kd,sd}).  In fact one
has here also (cf. \cite{By,ba,Bz}) $(\clubsuit)\,\,\Lambda\partial_{\Lambda}F
=-8\pi ib_1<Tr\phi^2>=-8\pi ib_1u$ so $\Lambda\partial_{\Lambda}F\sim
T_1\partial_1F$ in this situation.  Note from \cite{By,ba,Bz,ea,se}
$\tau=\partial^2_aF$ is dimensionless so $a(\partial_aF)_{\Lambda}+
\Lambda(\partial_{\Lambda}F)_a=2F$ when $F$ is thought of entirely in
terms of $a$ and $\Lambda$ variables; generally one writes,
when all $T_n$ variables vanish, $(\spadesuit)\,\,2F=(\Lambda
\partial_{\Lambda}+\sum m_j\partial/\partial m_j+\sum a_k\partial/\partial
a_k)F$ whereas $(\bullet)$ generalizes as $(\bullet\bullet)\,\,
2F=(\sum T_n\partial_n+\sum a_k\partial/\partial a_k+\sum m_j\partial/
\partial m_j)F$ without a scaling term.  
We note here a minus sign discrepency between \cite{ba} and \cite{se} for
example, along with a multiplier; thus in \cite{se} $\Lambda
\partial_{\Lambda}F=ib_1u/2\pi$ instead of $-8\pi ib_1u$ but 
\cite{ba} seems to fit better with \cite{ea,na} etc. so we follow this.
\\[3mm]\indent {\bf REMARK 1.2.}$\,\,$
It is tempting to suggest now that $T_1$ plays the role of a scaling
variable $\Lambda$ or cosmological term since $\Lambda\partial_{\Lambda}F=
T_1\partial_1F$, and some version of this
may have some validity
(cf. Section 6).  Certainly in the situation of Remark 1.1 for $T_1=c\Lambda$
one has $\Lambda\partial_{\Lambda}=T_1\partial_1$.
Generally one should look at
a finite number of
the $T_i$ as coordinates on the moduli space, just the
Casimirs $h_k$ are coordinates,
and their nature is that of coupling constants while their
role in $(\bullet\bullet)$ is to restore
the homogeneity of the prepotential, which
was destroyed by the nonvanishing beta function, and give a form for
$F$ compatible with the special geometry of $N=2$ supergravity
(cf. \cite{fg}).  
In \cite{ea} one claims that the $T_1$ variable can be identified as the
expectation value of the dilaton field 
and in \cite{ci,ya} $T_1=X$ is called a cosmological constant (see below
for more on this).  In any event $X\sim T_1$ can be allowed to play a
special role related to a puncture operator (but this
will change for more punctures).  
The problem of providing
physical interpretation of the 
other $T_n$ variables (corresponding to descendent fields)
seems to be related to describing
the gravity sector of $N=2$ supergravity.
Generally one can treat the $a_j$ variables as times in the spirit of
\cite{dc,Dy,ka,kc,ko} along with the $T_n$ and we will see that various
``moduli" seem to serve as coupling constants.
Renormalization theory (RT) usually works on
the space of coupling constants (or theories) so
the idea of connecting KP/Toda (or corresponding Whitham) times to RT
via beta functions
is not a priori unnatural. Recall also for $SU(2)$ SYM in \cite{ba} one uses 
beta functions 
\be
\beta(\tau)=(\Lambda\partial_{\Lambda}\tau)_u;\,\,
\beta^a(\tau)=(\Lambda\partial_{\Lambda}\tau)_a
\label{D}
\ee
where $\tau=\partial_a^2F=(\theta/2\pi)+(4i\pi/g^2)$ corresponds to
an effective coupling constant as well as the $\tau$ parameter of
an elliptic curve $E_u$ (cf. \cite{By,Bz} for important new contributions
to RG theory). 
Thus variations in $\tau$ measured by a
beta function have a fundamental geometric as well as physical meaning
(cf. \cite{Ka} for some refinements).
\\[3mm]\indent
Now let us give some general thoughts about integrability, Whitham,
renormalizability, etc. (partially extracted from sources as indicated).
In recent years the profusion of mathematical structures related to
integrability in various models of strings, quantum gravity, topological
field theory (TFT), conformal field theory (CFT), etc. made it possible
for a novice in physics to obtain the illusion of understanding a little bit
(even if restricted to 2-D toy models).  More recent progress in string theory
has led to M(matrix) theory, F theory, and a labyrinthine zoo of branes
from which the five basic string theories seem to emerge magically as special
situations.
Integrability is however still
visible amidst all this since, no matter how they emerge via
Calabi-Yau (CY) or brane wrappings,
the Riemann surfaces and
integrability directives of Seiberg-Witten (SW) theory
do in fact arise
(cf. \cite{de,da,gc,gd,ge,ia,kr,la,md,mf,me,mg}).
Thus there continues to be a fundamental role for
integrability and,
although this role may not have the unifying nature found in 2-D theories,
it represents an important substructure.  Another place
where integrability seems to
appear involves deformation
ideas, via the Whitham equations (renormalization may not be the correct
concept here).  
A good perspective on this does not
yet seem to have been written down, and we will only give some preliminary
remarks and a few
explicit formal calculations in this direction.
Let us also mention connections 
of Whitham equations to isomonodromic problems in the
spirit of \cite{hh,tc,td}. As
indicated in \cite{wa}, the phenomena described e.g. by 4-D Yang-Mills
(YM) equations are too complex to be described by an integrable system
and one does not expect quantum mechanics (QM) to be
integrable for a generic gauge group.  However, in view of the important
mathematical consequences concerning topology, algebra, and geometry
which have emerged from topics in QFT 
and string theory related to integrability, the concept needs no defense.
\\[3mm]\indent
In this direction
we extract first from \cite{tc} concerning the
origin of Whitham dynamics in SW theory
where it is stated 
(I believe correctly) that the derivation of the Whitham
dynamics looks very artificial, the least persuasive part being that it cannot
explain the origin of adiabatic deformation; it simply assumes that
deformation takes place.  The authors go on to say that origins are often
sought in (semi-classical) quantization of the classical integrable
systems which arise, which does not seem 
entirely satisfactory.  They further develop
an approach based on isomonodromic deformations which it is claimed might
eventually be absorbed into the idea of renormalization groups.  A key
feature here is the idea of multiscale analysis, and in any event it seems
to me that the idea of deformation should be regarded as fundamental.
In another direction a tantalizing idea comes from \cite{na} where it is
shown that the $N=2$ susy YM model could perhaps be interpreted as a
coupled system of two topological string models; the prepotential 
${\cal F}$ in fact plays the role of a free energy in a TFT or 
topological LG model.  This theme is discussed later.
Other comments about Whitham and renormalization
appear in \cite{gc,gd,ge,gg,md,mf} 
and we make a few comments based on these references as we go along.
Now renormalization group (RG) dynamics is governed by the action
of some vector field $d/d\,log\mu=\sum\beta_i(g)\partial/\partial g_i$
on the space of coupling constants $g_i(T)$ 
for example and Whitham dynamics gives
an example of some vector fields generated by slow time flows where 
coupling constant space is supplanted by a moduli space ${\cal U}$
of $u_{\alpha}$ where the $u_{\alpha}$ could be Casimirs, branch points,
coefficients of a LG superpotential, etc.  In the SW theory the
$u_{\alpha}\in {\cal U}$ are usually related to a spectral curve
$\Sigma\sim\Sigma_g$ (e.g. $\tau\in{\cal U}$ with $\beta=\Lambda
\partial_{\Lambda}\tau$); for Whitham times $T_i$ as moduli one
might look at $\mu\partial_{\mu}=\mu\sum (\partial T_i/\partial \mu)\partial_i
=\sum \beta_i\partial_i$, except that it is not clear how the $T_i$ depend on 
$\mu$.
In this spirit it is said (in a very unclear manner) that
Whitham is a generalization of RG equations
in the nonperturbative regime which still has the form of first order
differential equations in the coupling constants (which 
in turn correspond to the
coordinates in a moduli space ${\cal U}$ - recall also here that
$\tau=\tau(u)$ or $\tau=\tau(a)$ and $\tau=\partial^2{\cal F}/\partial a^2
=\oint_Bd\omega$ where $d\omega\sim$ normalized holomorphic differential).
Recall also that the normal variables of Whitham theory are certain
differentials $d\Omega_i$ on $\Sigma$ (or their coefficients in an
asymptotic expansion) or else Casimirs $h_k$ as in \cite{ia}.
The dependence of moduli $u_i,\,\,h_k$ etc. on flat (Whitham) times $T_n$
(for a finite set of $n$) is basically a coordinate change of moduli
however.
In any event dynamics on the moduli space ${\cal U}$ becomes important and
corresponds to dynamics in the space of coupling constants.
\\[3mm]\indent
The effective dynamics in the space of coupling constants (e.g. $\theta$
and $g^{-2}$
or $\tau$) replaces the original dynamics in space-time by a set of
Ward identities (low energy theorems) which normally have the form of
nonlinear differential equations for the effective action (which often
corresponds to a generalized tau function).
The parameter space here is
the spectral surface and vacua correspond to the family of spectral
surfaces.  This effective tau function induces a new (low energy
sector) dynamics on the space of moduli, identifying them as RG
slow dynamical variables of the theory.  Thus for hyperelliptic situations
the branch points $\Lambda_j$ can be moduli and label the vacua,
which correspond to finite zone solutions (of KP or Toda for example).
The Whitham dynamics on the $\Lambda_j$, or
LG coefficients $u_j$, or the Casimirs $h_k$ is induced by the
Riemann surface and the normal tau function via $\tau\to{\cal F}=\log\tau_{dKP}
\sim log\tau_{Whit}$
(the symbol $\tau$ is used for tau functions, as a modulus in $\beta=
\Lambda\partial_{\Lambda}\tau$, and later as a scaling variable in a base
curve for CM situations). 
In principle the Whitham
method of averaging over fast fluctuations required to produce
effective actions for slow variables, is said to play
the role of a nonperturbative
analogue of RG (this statement 
is much too vague and should be expanded).  
It seems that the $T_j$ are related to renormalized
KP/Toda times and the coupling constant $\tau
=\oint_Bd\omega$ above is emergent.
A theory (such as QCD) is asymptotically free if $g\to 0$ as 
$\Lambda\to\infty$ and $g\to\infty$ as $\Lambda\to 0$ ($g=0\sim$ a
free theory).  The behavior of a coupling constant is often described in
terms of its beta function $\beta=\Lambda\partial g/\partial\Lambda$ and 
an asymptotically free theory corresponds to $\beta<0$ for small $g$.
If $\beta\equiv 0$ the theory is scale invariant and coupling to matter
increases $\beta$.  In \cite{md} one suggests 
that the Whitham hydrodynamical type equations are generalizations of the
RG technique of perturbative theory but we question this.
\\[3mm]\indent
Next following \cite{ia},  
the problem of finding the low energy effective
action is formulated via ${\bf INPUT}:\,\,G=$ gauge group, $\tau=
(4\pi i/g^2)+(\theta/2\pi)\sim$ UV
bare coupling constant (which alternatively plays the role of a scaling
variable in \cite{dg} as in (\ref{ee}),
$m=$ mass scale, and $h_k=$ symmetry breaking
vev's $\to\,\,{\bf OUTPUT}:\,\,a_i(h)=$ background fields and ${\cal F}(a)=$
prepotential (and hence also $a_i^D=\partial{\cal F}/\partial a_i$ and
$\tau_{ij}=\partial{\cal F}/\partial a_i\partial a_j$).  The SW approach
was in effect to decompose this map via ${\bf (A)}:\,\,(G,\tau,m,h_k)\to
(\Sigma,d{\cal S}_{min})$ and ${\bf (B)}:\,\,(\Sigma,d{\cal S}_{min})\to
(a_i(h),{\cal F}(a))$, by formulas now very familiar.
Here one asks for 
$d{\cal S}$ as in \cite{ia} instead of some $d{\cal S}_{min}\sim\lambda_{SW}$
and a canonical formula is given.
The map ${\bf (A)}$ has no reference to 4 dimensions or to Yang-Mills (YM),
and represents something more primitive.  One looks for the map at the first
place where the group theory meets the algebraic geometry and this suggests
integrability theory.  Namely, {\bf (A)} possesses a description in terms of
1-D integrable models.  The only thing we need on the emergent integrable
system is its Lax operator $L(z)$ which is a $\tilde{g}^*$ valued matrix
function on
the phase space of the system and depends on the spectral parameter $z$ (on 
some base curve $E$, usually ${\bf P}^1$ or an elliptic curve $E(\tau)$).  Thus
${\bf (C)}:\,\,(G,\tau,m)\to L(z)$ and $\Sigma$ is determined via 
$det[t-L(z)]=0$
as a ramified cover of $E$.  The integrals of motion of the integrable
system are then identified with the 
moduli $h_k$.  The emphasis here is to determine
$L(z)$ and $\Sigma$ on the basis of group theory alone without recourse
to Hitchin varieties, geometric quantization, etc. (cf. \cite{de}) 
and the important concept of prepotential is still somewhat
unclear.  It is 
more fundamental than action and seems related to the fundamental role
that quasi-periodic trajectories (with ergodic properties) play in the 
transition from classical to quantum mechanics (one could run this back
to the Bohr-Sommerfeld atom).  
Why the theory of
quasiperiodic trajectories is expressible in terms of Hodge structure (special
geometry etc.) is apparently not understood.  Generally various theories
flow to the same universality class in the IR limit.  What the general
identification of effective actions with tau functions (i.e. with group
theory) teaches us is that these classes should be also representable
by some tau functions (not conventional ones defined via Lie group
terms).  In order to understand what the relevant objects are one considers
RG flow within some simple enough integrable system and discovers
that the relevant objects are quasiclassical tau functions or
prepotentials.  Another general
question here is how group theory (represented by generalized tau
functions) always flows to that of Hodge deformations
(represented by prepotentials). 

\section{RENORMALIZATION}
\renewcommand{\theequation}{2.\arabic{equation}}\setcounter{equation}{0}

This is a venerable subject and we make no attempt to survey it here
(for susy gauge
theories see e.g. \cite{By,ba,Bz,be,dz,dg,do,lf,ma,rd,
sk}).  In particular there are various geometrical ideas which can be introduced
in the space of theories $\equiv$ the space of coupling constants
(cf. here \cite{dy,di,dv,do,De,dp,le,oc,Sb,sa}).  We extract here
now mainly from \cite{De} where
it is argued that RG flow can be 
interpreted as a Hamiltonian vector flow on a phase space which consists
of the couplings of the theory and their conjugate ``momenta", which are
the vacuum expectation values of the corresponding composite operators.
For theories with massive couplings the identity operator plays a central
role and its associated coupling gives rise to a potential in the flow
equations.  The evolution of any quantity under RG flow can be obtained
from its Poisson bracket with the Hamiltonian.  Ward identities can be
represented as constants of motion which act as symmetry generators on the
phase space via the Poisson bracket structure.  
Consider a theory with $n-1$ couplings $g^a\,\,
(1\leq a\leq n-1)$ (sometimes one writes $g^a\sim g^a_R$ to denote
the renormalized coupling).  Let ${\cal M}$ be the space of couplings and
the beta functions $\beta^a(g)=dg^a/dt=\kappa(dg^a/d\kappa)\,\,(t=log(\kappa))$
constitute a vector field on ${\cal M}$ (assumed to be a differentiable
manifold or some sort - think of ${\bf R}^{n-1}$ for the moment).  The 
$2n-2$ dimensional with coordinates $(g^a,\beta^a)$ corresponds to the
tangent bundle $T({\cal M})$ and all the necessary information for computing
RG evolution is contained in the generating functional (or free energy)
$W(g,t)=\int w(g,t)d^Dx=-log(Z)$ where $w(g,t)$ is the free energy density
and $D\sim n-1$.  The phase space $T^*({\cal M})$ will have coordinates
$(g^a,\phi_a)$ where the $\phi_a$ are ``momenta" conjugate to the velocities
$\beta^a$ but no metric on $T({\cal M})$ is needed for the constructions.
A natural choice for the $\phi_a$ is the vev of the operator
associated with the coupling $g^a$, namely $\phi_a=(\partial w(g,t)/\partial
g^a)$.  For convenience one rescales all couplings by their canonical 
dimensions so that they become dimensionless in which case the $\phi_a$
are densities with mass dimension $D$ (one refers here to \cite{oc}
for guidelines).  One will produce a Hamiltonian which is linear in the
momenta and is minus the expectation value of the trace of the EM 
operator, written $H=-<T>$.  Despite the linearity $H$ is far from trivial.
The role of the identity operator is handled by introducing a coupling
$\Gamma$ (cosmological constant) whose conjugate momentum is the 
expectation value of the identity $I$
The corresponding beta function
is then $\beta^{\Gamma}(g,\Gamma)=-D\Gamma+U(g)
\,\,(\sim d\Gamma/dt)$ where $U(g)$ is an
analytic function, independent of $\Gamma$
(note $\beta^{\Gamma}=\partial g^{\Gamma}/\partial t=-D\Gamma +U$
is used and  $g^{\Gamma}\sim\Gamma$ - cf. \cite{bz,De}).
It results that $\phi_{\Gamma}=\kappa^D$ is also
a density and for massless theories one notes that $U(g)=0$  
(note $w\to w+\Gamma\kappa^D$).  For simplicity
one assumes first that the beta functions 
have no explicit dependence on $\kappa$
and only depend on $\kappa$ implicitly through $g^a(\kappa)$.
Then one shows (some details are indicated below
that the Hamiltonian $H(g,\phi)=\beta^a
(g)\phi_a+\beta^{\Gamma}(g,\Gamma)\phi_{\Gamma}$ governs the RG flow
of the $g^a$  and the expectation values $\phi_a$.  In fact the
RG evolution is given by ``Hamilton's equations"
\be
\frac{dg^a}{dt}=\left.\frac{\partial H}{\partial \phi_a}\right|_g;\,\,
\frac{d\phi_a}{dt}=-\left.\frac{\partial H}{\partial g^a}\right|_{\phi}
\label{NY}
\ee
The first equation is definitions while the second contains nontrivial
dynamics.
Now one extends the set $\{g^a\}$ to include $\Gamma$ and the
enhanced set $\{g^a,\Gamma\}$ will be denoted by $\{g^a\}$ again
with $1\leq a\leq n$ now; the corresponding space is denoted by
$\widehat{{\cal M}}$.  
One defines a Poisson bracket
\be
\{A,B\}=\frac{\partial A}{\partial g^a}\frac{\partial B}{\partial \phi_a}
-\frac{\partial A}{\partial \phi_a}\frac{\partial B}{\partial g^a}
\label{PY}
\ee
Evidently $\{g^a,\phi_b\}=\delta^b_a$ and the RG evolution for any function
on phase space is given by
\be
\frac{dA}{dt}=\left.\frac{\partial A}{\partial t}\right|_{g,\phi}+
\{A,H\}
\label{QY}
\ee
In particular when there is no explicit $\kappa$ dependence in the beta
functions, the Hamiltonian $H$ is a constant of motion (i.e. $dH/dt=0$).
One will also have an analogue of the Hamilton-Jacobi (HJ) equation
\be
\left.\frac{\partial w}{\partial t}\right|_g+H\left(g,\frac{\partial w}
{\partial g}\right)=0=\left.\frac{\partial 
w(g^a,t)}{\partial t}\right|_{g,\Gamma}+
\beta^a(g^a)\phi_a+\beta^{\Gamma}(g^a)\phi_{\Gamma}
\label{RY}
\ee
In fact writing $\beta^{\Gamma}=d\Gamma/dt=-D\Gamma+U(g^a)$ with
$w=w_R(g^a)+\Gamma\kappa^D$ one can express this via
\be
\left.\frac{\partial w_R(g^a,t)}{\partial t}\right|_g+
\beta(g^a)\phi_a+\kappa^DU(g^a)=0
\label{VY}
\ee
where $1\leq a\leq n-1$.  
\\[3mm]\indent
For the symplectic structure one begins with (\ref{VY}) in
the form 
\be
\phi_a(dg^a/dt)+(\partial w/\partial t)=0
\label{WWY}
\ee
with $1\leq a\leq n$
and $w=w_R+\kappa^D\Gamma$.  To emphasize the analogy with classical
mechanics one defines a function $H$ via $H=-\partial w/\partial t$ so that
the basic RG equation involves $H(g,\phi)=\beta^a(g)\phi_a+
\beta^{\Gamma}\phi_{\Gamma}$.  It is assumed that the beta functions
have no explicit $\kappa$ dependence so that $H$ has no explicit
$t$ dependence.  One treats the $\phi_a$ as independent variables and 
after the theory has been solved one uses $\phi_a=\partial_aw$
(see \cite{De} for further details).
In any event
from $H=\beta^a(g)\phi_a+\beta^{\Gamma}\phi_{\Gamma}$ one has
\be
\frac{d\phi_a}{dt}=-\frac{\partial}{\partial g^a}\left(\kappa^DU(g)\right)
-\left(\frac{\partial\beta^b}{\partial g^a}\right)\phi_b\Rightarrow
\frac{d\phi_a}{dt}+(\partial_a\beta^b)\phi_b=-\kappa^D\partial_aU
\label{cy}
\ee
This is the RG for RG evolution of the vev's of the basic operators of
the theory.  There is a parallel with Newton's second law in that the
matrix of anomalous dimensions $\partial_a\beta^b$ appears as a 
pseudo-force (Coriolis force) and $U(g)$ is a potential.  For massless
theories $U$ vanishes so this is analogous to free particle motion.
Once the theory is solved
(\ref{cy}) becomes
\be
\left.\frac{\partial\phi_a}{\partial t}\right|_g+\beta^b\partial_b\phi_a
+(\partial_a\beta^b)\phi_b=-\kappa^D\partial_aU
\label{gy}
\ee
This is a version of the RG equation for the vev's, including the
anomalous dimensions and the inhomogeneous term $-\kappa^D\partial_aU$
which arises due to masses.  Another way of expressing this involves
the Lie derivative $L_{\vec{\beta}}\phi$ where $\vec{\beta}\sim\beta^a
(\partial/\partial g^a)$ (cf. \cite{De}).
The analogy with classical mechanics goes still further via
$H(g,\phi)+w_t=0$ to a HJ equation ($\phi_a=(\partial w/\partial g^a)$ when
the theory is solved)
\be
\left.\frac{\partial w}{\partial t}\right|_g+H\left(g,\frac{\partial w}
{\partial g}\right)=0
\label{jy}
\ee
\indent
All this structure suggests a reformulation of the RG using phase
space variables.  A quantity $A$ is considered
as a function of $(g^a,\phi_a)$ and possibly the RN point $t=log(\kappa)$
with evolution given via
\be
\frac{dA}{dt}=\frac{\partial A}{\partial g^a}\frac{\partial H}{\partial\phi_a}
-\frac{\partial A}{\partial\phi_a}\frac{\partial H}{\partial g^a}+
\left.\frac{\partial A}{\partial t}\right|_{g,\phi}=\{A,H\}+
\left.\frac{\partial A}{\partial t}\right|_{g,\phi}
\label{ky}
\ee
Since there is no explicit $\kappa$ dependence in $H$ one has
$dH/dt=0$ (provided there is no explicit $\kappa$ dependence in the
beta functions).  Formulas for RG evolution of $N$ point Green's
functions are developed 
in \cite{De} along with a rich supply of remarks.  In particular
one notes that $H=\beta^a\phi_a+\beta^{\Gamma}\phi_{\Gamma}$ has a
simple interpretation.  The right side of this equation is the negative
of the usual definition of the vev of the trace of the EM tensor,
$H=-<T>$ and it should be no surprise that $<T>=(\partial w/\partial t)|_g$ 
since varying $t$ with the couplings fixed is completely equivalent to
a conformal rescaling of the metric.  The derivative $(\partial/\partial t)|_g$
acting on $w$ simply pulls down the action from the exponent and then varies
the metric leading to $<T_{\mu}^{\mu}>$.  Thus the entire RG evolution is
governed solely by $<T>$ (cf. \cite{De} for more on this).  
At fixed points of the RG flow (conformal
field theories) the Hamiltonian vanishes because the beta functions do.
One can ask what is the special ingredient of the RG flow which allows
it to be written in Hamiltonian form.  The crucial fact is $dw/dt=0$
which means that the RG is a symmetry.  
\\[3mm]\indent
The background here can also be made clearer following 
\cite{do,oc}.  Thus consider
\be
Z(g)=\int{\cal D}\phi\,exp(-S(\phi));\,\,W(g)=-log\,Z(g)=\int wd^Dx
\label{64}
\ee
where $S\sim$ action and e.g. $w=W/V$ for $V=\int d^Dx$.  Then
\be
1=\int{\cal D}\phi e^{W-S(\phi)}\Rightarrow dW=<dS>=\int{\cal D}\phi
dS(\phi)e^{W-S(\phi)}
\label{65}
\ee
where $dW=\partial_aWdg^a$ and $dS=\partial_aSdg^a$.  If the action is
linear in the couplings, e.g. $S\sim\int d^Dx\sum g^a\hat{\Phi}_a$, then
$(\bullet\bullet\bullet)\,\,\partial_aS\sim\int d^Dx\hat{\Phi}_a$;
(\ref{65}) and $(\bullet\bullet\bullet)$ can then be referred to as an
action principle.  A metric advocated in \cite{oc}
involves a line element
\be
ds^2=<(dS-dW)\otimes (dS-dW)>
\label{66}
\ee
on the $g^a$ parameter space.  Then one divides this by $V$ and uses
densities with (formally)
\be
\tilde{\Phi}_a=\hat{\Phi}_a-<\hat{\Phi}>;\,\,G_{ab}=\int d^Dx
<\tilde{\Phi}_a(x)\tilde{\Phi}_b(0)>
\label{67}
\ee
This is formally acceptable as a metric.  Setting $w=W/V$ as above
one has from (\ref{65}) $\partial_aw=(1/V)<\partial_aS>$ and hence
\be
\partial_a\partial_bw=\frac{1}{V}\left\{<\partial_a\partial_bS>-
<\partial_aS\partial_bS>+<\partial_aS><\partial_bS>\right\}
\label{68}
\ee
which implies
$G_{ab}=(1/V)<\partial_a\partial_bS> -\partial_a\partial_bw$.
One checks that this is covariant under general coordinate transformations.
When $S$ is linear in the couplings it implies $(\bullet\clubsuit)\,\,
G_{ab}=-\partial_a\partial_bw$, and one must work then with linear
coordinate transformations (or with Legendre transformed variables - cf.
\cite{do}).
\\[3mm]\indent
Finally a word on the situation when $\beta^a=\beta^a(g,t)$ for example.
Then take $t$ as an additional coupling and enlarge the space to an 
$n+1$ dimensional $\hat{{\cal M}}=\{g^a,\Gamma,t\}$.  the momentum
conjugate to $t$ is $-H$ where $\phi_t=\partial_tw=-H(g,\phi,t)$ and note
$\beta^t=1$.  The Hamiltonian on $T^*(\hat{{\cal M}})$ is $H_E=
\sum\beta^a(g,t)\phi_a+\beta^{\Gamma}\phi_{\Gamma}+\phi_t$ and a
new evolution parameter $\tau$ is introduced with
\be
\frac{d\phi_t}{d\tau}=-\left.\frac{\partial H_E}{\partial t}\right|_{g,\phi}=
-(\partial_t\beta^a)\phi_a
\label{100}
\ee
over all $a\sim (a,\Gamma)$.  When the theory is solved $H_E=0$ which is
the HJ equation (\ref{jy}) and for $\tau=t$ (\ref{100}) becomes
\be
\frac{dH}{dt}=\frac{\partial H}{\partial t}=\sum
\beta^a_t\phi_a+\beta_t^{\Gamma}\phi_{\Gamma}
\label{101}
\ee
Thus the $t$ dependent Hamiltonian $H$ is not RG invariant.

\section{RG AND SUSY GAUGE THEORIES}
\renewcommand{\theequation}{3.\arabic{equation}}\setcounter{equation}{0}

A fascinating study of RG in a Whitham framework appears in 
\cite{dz,dg} (cf. also \cite{Mq})
and we refer to \cite{bh,cc,cg,ka,kc,ko,na}
for background; here we only sketch the framework and summarize
a few results from \cite{dz,dg}.
First consider $N=2,\,\,SU(N_c)$ gauge
theories with $N_f$ quark flavors and $N_f<2N_c$.  There are $N_f$
hypermultiplets of bare masses $m_j$ and the $N=2$ chiral multiplet
contains a complex scalar field $\phi$ in the adjoint representation.
The classical moduli space of vacua is $N_c-1$ dimensional and can be
parametrized by eigenvalues $\bar{a}_k$ of $\phi$ where $\sum\bar{a}_k=0$.  
For generic $\bar{a}_k$ the $SU(N_c)$ gauge symmetry is broken to
$U(1)^{N_c-1}$ and in the $N=1$ formalism the Wilson effective 
Lagrangian of the quantum theory to leading order in the low momentum
expansion is 
\be
L=\frac{1}{4\pi}\Im\left[\int d^4\theta
\frac{\partial{\cal F}(A)}{\partial A^i}\bar
{A}^i+\frac{1}{2}\int d^2\theta\frac{\partial^2{\cal F}(A)}{\partial A^i
\partial A^j}W^iW^j\right]
\label{O}
\ee
Here the $A^i$ are $N=1$ chiral superfields whose scalar components
correspond to the $\bar{a}_i$.  For $SU(N_c)$ theories with $N_f<2N_c$
one should have ${\cal F}$ expressed in terms of a classical
prepotential plus a one loop term plus instanton contributions via
\be
{\cal F}(A)=\frac{1}{2\pi i}(2N_c-N_f)\sum_1^{N_c}A^2_i+\sum_1^{\infty}
{\cal F}_d(A_k)\Lambda^{(2N_c-N_f)d}-
\label{P}
\ee
$$-\frac{1}{8\pi i}\left(\sum (A_k-A_n)^2log\frac{(A_k-A_n)^2}{\Lambda^2}
-\sum_1^{N_c}\sum_1^{N_f}(A_k+m_j)^2log\frac{(A_k+m_j)^2}{\Lambda^2}\right)$$
The spectral curves will have the form (cf. \cite{dz,dg,ka,ko})
\be
y^2=A^2(x)-B(x);\,\,d\lambda=\frac{x}{y}\left(A'-\frac{1}{2}(A-y)
\frac{B'}{B}\right)dx
\label{Q}
\ee
(there is no relation between the A's in (\ref{P}) and (\ref{Q})).
Specifically, let $\Lambda$ be the dynamically generated scale of the theory
with $\bar{s}_i,\,\,0\leq i\leq N_c$ and $t_p(m),\,\,1\leq p\leq N_f$
the i-th and p-th symmetric polynomials in $\bar{a}_k$ and $m_j$
respectively, i.e.
\be
\bar{s}_i=(-1)^i\sum_{k_1<\cdots <k_i}\bar{a}_{k_1}\cdots\bar{a}_{k_i};\,\,
t_p(m)=\sum_{j_1<\cdots <j_p}m_{j_1}\cdots m_{j_p}
\label{R}
\ee
The polynomials $A$ and $B$ are given by
\be
A(x)=C(x)+\frac{\Lambda^{2N_c-N_f}}{4}T(x);\,\,B(x)=\Lambda^{2N_c-N_f}
\prod_1^{N_f}(x+m_j);
\label{T}
\ee
$$C(x)=\prod_1^{N_c}(x-\bar{a}_k)=x^{N_c}+\sum_2^{N_c}\bar{s}_ix^{N_c-i}$$
where $T(x)$ is a certain polynomial.
One can absorb the $T(x)$ dependence in a
redefinition of the classical order parameters $\bar{a}_k$, since
the addition of $T(x)$ just modifies the bare parameters $\bar{s}_i$ in
(\ref{T}) to parameters $\tilde{s}_i$, via
\be
A(x)=x^{N_c}+\sum_2^{N_c}\tilde{s}_ix^{N_c-i}=\prod_1^{N_c}
(x-\tilde{a}_k);\,\,\tilde{s}_i=\bar{s}_i+\frac{1}{4}\Lambda^{2N_c-N_f}t_i
\label{V}
\ee
The Riemann surface $\Sigma$ of (\ref{Q}) in this context is a double cover of
the complex plane with branch points $x_k^{\pm},\,\,1\leq k\leq N_c$
defined via $A(x_k^{\pm})^2-B(x_k^{\pm})=0$.  For $\bar{\Lambda}=
\Lambda^{N_c-N_f/2}$ small the $x_k^{\pm}$ are just perturbations of
the $\bar{a}_k$.  One can view $\Sigma$  as two copies of {\bf C}, cut 
and joined along slits from $x_k^{-}$ to $x_k^{+}$, with canonical
homology basis $(A_k,B_k)\,\,(2\leq k\leq N_c)$, where $A_k$ is a simple
contour enclosing the slit from $x_k^{-}$ to $x_k^{+}$ and $B_k$
is a curve going from $x_k^{-}$ to $x_1^{-}$ on each sheet.
The renormalized order parameters are given by
\be
2\pi ia_k=\oint_{A_k}d\lambda=\oint_{A_k}dx
\frac{x\left(\frac{A'}{A}-\frac{B'}{2B}\right)}{\sqrt{1-\frac{B}{A^2}}}
\label{W}
\ee
\indent
One can now introduce the more or less standard machinery of 
Baker-Akhiezer (BA) functions, tau functions, etc. for a RS with
punctures leading to dispersionless theory and the Whitham equations.
We mainly refer to \cite{Aa,bh,cc,cg,ch,ci,cj,dz,dg,dc,fb,gb,ka,
kb,kc,ko,na,ta} for all that since we want to concentrate on other
matters in this paper. 
One takes $\lambda_{SW}\sim d\lambda=QdE=dS$ for suitable meromorphic
differentials $dQ$ and $dE$.  There will be Whitham times $T_A$, dual
times $T_A^D$, and associated differentials $d\Omega_A$ such that
$\partial_Ad\Omega_B=\partial_Bd\Omega_A$ and $\partial_AdS=d\Omega_A$
where $\partial_A\sim\partial/\partial T_A$ 
leading to $Q=\partial S/\partial E$. 
The Whitham tau function is $\tau=exp({\cal F})$ where ${\cal F}$ coincides
with the prepotential in the SW situation (modulo $\pm2\pi i$ factors
which come and go - not to worry here).  One has a large moduli space for
RS $\Sigma_g$ with punctures and prescribed pole behavior of $dE$ and $dQ$ at
the punctures.  One specifies a foliation by level sets of certain moduli
and works on a leaf of this foliation.  Relations to TFT and topological
LG theories are spelled out and one has WDVV equations etc. (cf.
\cite{By,ba,cw,dw,dc} - more on this later).
\\[3mm]\indent
For $N=2$ susy YM theories in 4-D with gauge group $G$ the YM gauge
field $A=A_{\mu} dx^{\mu}$ is imbedded in an $N=2$ gauge multiplet consisting
of $A$, left and right spinors $\lambda_L$ and $\lambda_R$, and a complex
scalar field $\phi$, with all fields in the adjoint representation of $G$.
The requirement of $N=2$ susy and renormalizability fixes uniquely the
action
\be
I=\int_{M^4}d^4xTr\left[\frac{1}{4g^2}F\wedge F^*+\frac{\theta}{8\pi^2}
F\wedge F +D\phi^{\dagger}\wedge *D\phi+[\phi,\phi^{\dagger}]^2\right]
+\,\,fermions
\label{a}
\ee
Here $g$ is the coupling constant and $\theta$ is the instanton angle.
The classical vacua involve $[\phi,\phi^{\dagger}]=0$ so $\phi$ lies in
the Cartan subalgebra and one writes
\be
\phi=\left(
\begin{array}{cccc}
\bar{a}_1 & {} & {} & {}\\
{} & \bar{a}_2 & {} & {}\\
{} & \cdots & \cdots &{}\\
{} & {} & {} & \bar{a}_{N_c}
\end{array}\right);\,\,\sum_1^{N_c}\bar{a}_k=0
\label{b}
\ee
Generically $\bar{a}_j\not= \bar{a}_k$ and the gauge group is spontaneously
broken to $U(1)^{N_c-1}$.  At the quantum level one expects then that the
space of inequivalent vacua will be parametrized by $N_c-1$ parameters
$a_k$ with $\sum_1^{N_c}a_k=0$ (thought of as renormalizations of the
$\bar{a}_k$).  Each vacuum corresponds to a theory of $N_c-1$ interacting
$U(1)$ gauge fields $A_j$ (copies of electromagnetism (EM)).  Since
$N=2$ susy remains unbroken each gauge field $A_j$ is part of an $N=2$
susy $U(1)$ gauge multiplet $(A_j,\lambda^j_L,\lambda^j_R,\phi_j)$
all in the adjoint representation of $U(1)$.  To leading order in
the low momentum expansion one has an effective action
$$
I_{eff}=\frac{1}{8\pi}\int_{M^4}d^4x\left[(\Im\tau^{jk})F_j\wedge *F_k
+(\Re\tau^{jk})F_j\wedge F_k+d\phi_i^{\dagger}\wedge d\phi_i^D\right]
+\,\,fermions;$$
\be
\tau^{jk}=\frac{\partial^2{\cal F}}{\partial a_j\partial a_k};\,\,\phi_j^D=
\frac{\partial{\cal F}}{\partial a_j}(\phi)
\label{c}
\ee
One thinks here of ${\cal F}(a,\Lambda)$ where $\Lambda$ is the renormalization
scale.  In order to have positive kinetic energy one posits
$(\clubsuit\clubsuit)\,\,\Im[\partial^2{\cal F}/\partial a_j
\partial a_k]>0$ so ${\cal F}$ defines a K\"ahler metric on the quantum
moduli space via $(\spadesuit\spadesuit)\,\,ds^2=\sum\Im[\partial^2
{\cal F}/\partial a_j\partial a_k]da_j\bar{da_k}$.  At weak coupling
$\Lambda<< 1,\,\,{\cal F}$ can be evaluated in perturbation theory
and for pure $SU(N_c)$ YM one has
\be
{\cal F}(a,\Lambda)=\frac{2N_c}{2\pi i}\sum_1^{N_c}a^2_k-\frac{1}{8\pi i}
\sum_1^{N_c}(a_k-a_j)^2log\frac{(a_k-a_j)^2}{\Lambda^2}+
\sum_1^{\infty}{\cal F}_d\Lambda^{2N_cd}
\label{d}
\ee
In the presence of $N_f$ hypermultiplets in the fundamental representation
of bare masses $m_i\,\,(1\leq i\leq N_f)$ there will be an additional
term in the one loop correction for the $SU(N_c)$ theory, namely
\be
\sum_1^{N_c}\sum_1^{N_f}(a_k+m_j)^2log\frac{(a_k+m_j)^2}{\Lambda^2}
\label{e}
\ee
\indent
Then the SW ansatz requires that for each $\Lambda$ the quantum moduli
space should parametrize a family of RS $\Sigma(a,\Lambda)$ of genus
$N_c-1$ with a meromorphic one form $\lambda_{SW}$ on each $\Sigma$
determining ${\cal F}$ via the periods
\be
a_k=\frac{1}{2\pi i}\oint_{A_k}\lambda_{SW};\,\,a^D_k=\frac{1}{2\pi i}
\oint_{B_k}\lambda_{SW};\,\,\frac{\partial{\cal F}}{\partial a_k}=a_k^D
\label{f}
\ee
Now one will identify
$\lambda_{SW}$ with $d\lambda=QdE$.
For $SU(N_c)$ theories with $N_f<2N_c$
hypermultiplets having bare masses $m_i$ the spectral curves are given
by the leaf $(\Sigma,E,Q)$ with the following properties:  ${\bf (A)}\,\,
dE$ has simple poles at points $P_{\pm},\,P_i$ with residues $-N_c,\,
N_c-N_f$, and $1\,\,(1\leq i\leq N_f)$.  Its periods around homology
cycles are multiples of $2\pi i.\,\,
{\bf (B)}\,\,Q$ is a meromorphic function with simple poles only at
$P_{\pm}.\,\,
{\bf (C)}\,\,$.  The other parameters of the leaf are determined by the
following normalizations of $d\lambda=QdE$
\be
Res_{P_i}(d\lambda)=-m_i;\,\,Res_{P_{+}}(zd\lambda)=-N_c2^{-1/N_c};
\label{ff}
\ee
$$Res_{P_{-}}(zd\lambda)=(N_c-N_f)\left(\frac{\Lambda^{2N_c-N_f}}{2}\right)^
{1/(N_c-N_f)};\,\,Res_{P_{+}}(d\lambda)=0$$
These conditions imply that $\Sigma$ is hyperelliptic
and has an equation of the form
\be
y^2=\prod_1^{N_c}(Q-\bar{a}_k)^2-\Lambda^{2N_c-N_f}\prod_1^{N_f}(Q+m_j)
\equiv A(Q)^2-B(Q)
\label{g}
\ee
(cf. here (\ref{Q})).  Strictly speaking the parameters $\bar{a}_k$ agree
with classical vacua only when $N_c< N_f$.  For $N_f\geq N_c$ there are
$O(\Lambda)$ corrections which can be absorbed in a reparametrization
leaving ${\cal F}$ invariant; hence we identify the $\bar{a}_k$ of
(\ref{b}) and (\ref{g}).  If one represents the RS (\ref{g}) by a two
sheeted covering of the complex plane then $Q$ is just the coordinate
in each sheet while $(\clubsuit\clubsuit\clubsuit)\,\,
E=log(y+A(Q))$.  
The points $P_{\pm}$ are points at
infinity with the two possible sign choices $\pm$ for $y=\pm\sqrt
{A^2-B}$.  The constructions proceed as in (\ref{Q}) - (\ref{W})
leading to 
$(\spadesuit\spadesuit\spadesuit)\,\,a_k=\bar{a}_k+O(\Lambda^{N_c})$.  The
prepotential then satisfies ($z$ is a local coordinate)
\be
\sum_1^{N_c}a_j\frac{\partial{\cal F}}{\partial a_j}+\sum_1^{N_f}m_j\frac
{\partial{\cal F}}{\partial m_j}-2{\cal F}= {\cal D}{\cal F}=
\label{h}
\ee
$$=-\frac{1}{2\pi i}
\left[Res_{P_{+}}(zd\lambda)Res_{P_{+}}(z^{-1}d\lambda)
+Res_{P_{-}}(zd\lambda)Res_{P_{-}}(z^{-1}d\lambda)\right]$$
It is known that the right side of
(\ref{h}) is a modular form (cf. \cite{bc,ba,be}) and one arrives at
\be
{\cal D}{\cal F}=\frac{1}{4\pi i}(N_f-2N_c)\sum_1^{N_c}\tilde{a}_k^2
\label{q}
\ee
\indent {\bf REMARK 3.1.}$\,\,$ 
Referring to (\ref{Q}) - (\ref{W}) where $y^2=A^2-B$ we compare to \cite
{na} and \cite{ea}.  Thus in \cite{na} $y^2=P^2-\Lambda^{2N}$ with 
$P=x^N+\sum_0^{N-2}u_{N-k}x^k$ corresponding to (\ref{T}) with
$N_f=0$ and in \cite{ea} one has $N_f< N_c$ with $y^2=C(x)^2-
\Lambda^{2N_c-N_f}G(x)$ where
\be
C=x^{N_c}-\sum_2^{N_c}u_ix^{N_c-i};\,\,G=\prod_1^{N_f}(x+m_j)\,\,(u_2=u);
\label{k}
\ee
$$\lambda\sim\frac{dz}{2\pi i}\left[\left(\frac{N_f}{2}-N_c\right)z^{-2}
-\frac{1}{2}\sum_1^{N_f}m_jz^{-1}+\left(-2u+\frac{1}{2}\sum_1^{N_f}
m_j^2\right)\right]$$
(the latter on the $P_{+}$ sheet).  This implies in the context of \cite{ea}
\be
T_1=\frac{1}{2\pi i}\left(N_c-\frac{N_f}{2}\right);\,\,T_0=-\frac{1}{4\pi i}
\sum_1^{N_f}m_j;\,\,\frac{\partial{\cal F}}
{\partial T_1}=2u-\frac{1}{2}\sum_1^{N_f}m_j^2
\label{l}
\ee
In the massless limit $\sum a_j(\partial{\cal F}/
\partial a_j)-2{\cal F}=8\pi ib_1u=-T_1\partial_1{\cal F}$ (cf.
Remark 1.1) and here $b_1=(2N_c-N_f)/16\pi^2$.
Thus apparently two times are needed to adjust ${\cal F}$ in this
situation and this casts some doubt on the eventual identification
of $\Lambda$ (or $log(\Lambda$)) with any one $T_j$ parameter.
\\[3mm]\indent
Now in \cite{dg} one begins with an elliptic CM system
\be
p_i=\dot{x}_i;\,\,\dot{p}_i=m^2\sum_{j\not= i}{\cal P}'(x_i-x_j);\,\,
1\leq i,j\leq N
\label{aa}
\ee
This admits a Lax representation $\dot{L}=[M,L]$ with $N\times N$ matrix
entries (cf. below) and a spectral parameter $z$ living on a torus
$\Sigma$.  The complex modulus of the torus is 
$\tau=(\theta/2\pi)+(4\pi i/e^2)$ and the spectral curve is given by
$R(k,z)=det(kI-L(z))=0$ with SW differential given via $d\lambda=
kdz$ (notation
may vary at times).  One is dealing here with the adjoint representation
where the match between 4-D gauge theory and 2-D integrable models was
originally found by indirect arguments and the order parameters are
difficult to recognize.  One will seek a single monic polynomial
$H(k)=\prod_1^N(k-k_i)$ whose zeros $k_i$ are essentially the classical
order parameters of the guage theory.  More precisely one sets
\be
f(k,z)=R\left(k-m\partial_z\theta_1\left[\frac{z}{2\omega_1}
\mid \tau\right],
z\right)
\label{bb}
\ee
then the elliptic CM spectral curves are characterized by
\be
f(k,z)=\frac{1}{\theta_1(\frac{z}{2\omega_1}|\tau)}\theta_1\left(
\frac{1}{2\omega_1}\left\{z-m\frac{\partial}{\partial k}\right\}
\mid\tau\right)
H(k)
\label{cc}
\ee
and the classical order parameters are given via
\be
k_i-\frac{1}{2}m=lim_{q\to0}\oint_{A_j}d\lambda;\,\,\,q=e^{2\pi i\tau}
\label{dd}
\ee
Then one finds formulas for the prepotential ${\cal F}$ and arrives
eventually at the lovely formula
\be
\frac{\partial{\cal F}}{\partial\tau}=\frac{1}{4\pi i}\sum_1^N\oint_{A_j}k^2dz
\label{ee}
\ee
which is a RG equation connecting the RG beta function of the 4-D gauge
theory (with abuse of notation)
to the Hamiltonian of the 2-D CM system.  
Note that the ``coupling constant" $\tau$ of the base curve is playing the
role of a scaling variable here as indicated below (cf. (\ref{ao})) and
$\partial{\cal F}/\partial\tau$ officially should not perhaps be called
a beta function unless ${\cal F}$ can be thought of as a coupling constant.
When the full
hypermulitplet is decoupled one obtains the pure $N=2$ susy $SU(N)$ gauge
theory and (\ref{ee}) reduces accordingly (see below).  
At first passage here we will
largely ignore the connections to Hitchin systems and the approach in
\cite{da} but this is made explicit in \cite{dg}.
We emphasize that $\tau$ is the parameter for the base
curve $E(\tau)$ here and its role in (\ref{ee}) is that of a scaling
variable in RG theory.  Thus e.g. $\tau=(1/2\pi i)log(q)$ and $q\partial_q=
\partial/\partial(log(q))$ means $q\partial_q{\cal F}=-(1/8\pi^2)\sum_1^N\oint_
{A_j}k^2dz$ (cf. also \cite{ia} here).
\\[3mm]\indent
The independent parameters in the $N=2$ susy $SU(N)$ gauge theory are
the complex gauge coupling $\tau$, the hypermultiplet mass parameter
$m$, and the quantum order parameters $a_i,\,\,1\leq i\leq N$ (or
equivalently the classical order parameters $k_i$).  In \cite{dg} one
now considers various decoupling limits of the $N=2$ theory with a
massive adjoint hypermultiplet.  The case of most interest here
(to me at least) involves $\tau\to i\infty$ (so $q\to 0$) and 
$m\to \infty$ while the parameters $a_i$ and $\Lambda$ remain finite
(here $({\bf Z})\,\,\Lambda^{2N}=(-1)^Nm^{2N}q$); it is equivalent to keep the
classical order parameters $k_i$ fixed.  Upon scaling $w$ in such a
way that $t$ defined by $w=t(-m)^{-N}$ is kept fixed where
$H(k)-t-(\Lambda^{2N}/t)=0$, the spectral curve converges to the
SW curve of the pure theory. Further
the SW differential follows directly
from the same change of variable $z=log(w)$ to $t$ yielding
$d\lambda=kd\,log(t)$.  Finally the so-called RG equation (\ref{ee}) reduces
to a RG equation for the pure theory as in \cite{dz}.  Namely, the
sum over the $A_j$ cycles in (\ref{ee}) may be deformed into a single contour
encircling all the $A_j$ which in turn may be deformed into a contour 
around $\infty$.  Upon defining $s_2$ via $H(k)=k^N+s_2k^{N-2}+O(k^{N-3})$
and using ({\bf Z}) one obtains
\be
\frac{\partial{\cal F}}{\partial\,log(\Lambda)}=-\left(\frac{N}{\pi i}
\right)s_2
\label{ao}
\ee
in agreement with \cite{dz} (cf. (\ref{q})).
We note that the Whitham times $T_n$ are suppressed in (\ref{q}),
(\ref{ee}), and
(\ref{ao}) since one was working on certain leaves of a foliation;
they would be eventually used to restore the homogeneity
of ${\cal F}$.  These equations have a flavor reminiscent of the
Zamolodchikov C theorem 
(cf. \cite{By,Bz,dy,gh,lf,mu,zc,zb}) and we will return to this
later; the clarification of \cite{By,Bz} seems definitive.

\section{ADE AND LG APPROACH}
\renewcommand{\theequation}{4.\arabic{equation}}\setcounter{equation}{0}

Connections of TFT, ADE, and LG models abound (cf. \cite{Aa,ci,dw,dc,
kc,ko,ta,ya}) and for $N=2$ susy YM we go to \cite{iz} 
(cf. also \cite{bl,eb}).  Thus one evaluates integrals $a_i=\oint_{A_i}
\lambda_{SW}$ and $a_i^D=\oint_{B_i}\lambda_{SW}$ using Picard-Fuchs
(PF) equations.  One considers $P_R(u,x_i)=det(x-\Phi_R)$ where 
$R\sim$ an irreducible representation of $G$ and $\Phi_R$ is a representation
matrix.  Let $u_i\,\,(1\leq i\leq r)$ be Casimirs built from $\Phi_R$ of
degree $e_i+1$ where $e_i$ is the $i^{th}$ exponent of $G$ (see below).
In particular $u_1\sim$ quadratic Casimir and $u_r\sim$ top Casimir of 
degree $h$ where $h$ is the dual Coxeter number of $G$ ($h=r+1$ for 
$A_r$).  The quantum SW curve is then
\be
\tilde{P}_R(x,z,u_i)\equiv P_R\left(x,u_i+\delta_{i,r}\left[z+
\frac{\mu^2}{z}\right]\right)=0
\label{28}
\ee
where $\mu^2=\Lambda^{2h}$ with $\Lambda\sim$ the dynamical scale and the
$u_i$ are considered as gauge invariant moduli parameters in the Coulomb
branch.  This curve is viewed as a multisheeted foliation $x(z)$ over
${\bf CP^1}$ and the SW differential is $\lambda_{SW}=x(dz/z)$.  The physics
of $N=2$ YM is described by a complex $rank(G)$ dimensional subvariety of
the Jacobian which is a special Prym variety (cf. \cite{da,mg}).  Now
one writes (\ref{28}) in the form
\be
z+\frac{\mu^2}{z}+u_r=\tilde{W}_G^R(x,u_1,\cdots,u_{r-1})
\label{29}
\ee
For the fundamental representations of $A_r$ and $D_r$ one has
\be
\tilde{W}_{A_r}^{r+1}=x^{r+1}-u_1x^r-\cdots-u_{r-1}x;\,\,
\label{30}
\ee
$$\tilde{W}_{D_r}^{2r}=x^{2r-2}-u_1x^{2r-4}-\cdots -u_{r-2}x^2-
\frac{u_{r-1}^2}{x^2}$$
and setting ${\bf (SP)}\,\,W_G^R(x,u_1,\cdots,u_r)=\tilde{W}^R_G(x,
u_1,\cdots,u_{r-1})-u_r$ it follows that $W^{r+1}_{A_r}$ and $W^{2r}_{D_r}$
are the fundamental LG superpotentials for $A_r$ and $D_r$ type topological
minimal models (cf. also \cite{dw,dc,eb}).
The $u_i$ can be thought of as coordinates on the space of TFT and the
presentation in \cite{dw} for $A_{n-1}$ is 
somewhat clearer (cf. Remark 5.1).
We will concentrate on $A_r$ but $D_r$ and other groups are discussed in
\cite{iz}.
Now in 2-D TFT of LG type $A_r$ with superpotential {\bf (SP)} the flat
time coordinates for the moduli space are given via
\be
T_i=c_i\oint dx\,W^R_G(x,t)^{e_i/h}\,\,\,(i=1,\cdots,r)
\label{31}
\ee
These are residue calculations for Whitham type times $T_i$ which
will be polynomials in the $u_j$ (the normalization
constants $c_i$ are indicated in \cite{iz}.  One defines primary fields
\be
\phi_i^R(x)=\frac{\partial W^R_G(x,u)}{\partial T_i}\,\,\,(i=1,\cdots,r)
\label{32}
\ee
where $\phi_r^R=1$ is the identity $\sim$ puncture operator.  The one
point functions of the gravitational descendents $\sigma_n(\phi_i^R)$
(cf. \cite{Aa,dw,dc}) are evaluated via
\be
<\sigma_n(\phi_i^R)>=b_{n,i}\sum_1^r\eta_{ij}\oint W_G^R(x,u)^
{(e_j/h)+n+1}\,\,\,(n=0,1,\cdots)
\label{33}
\ee
for certain constants $b_{n,i}$ (cf. \cite{iz} for details).  The
topological metric $\eta_{ij}$ is given by
\be
\eta_{ij}=<\phi_i^R\phi_j^RP>=b_{0,r}\frac{\partial^2}{\partial T_i
\partial T_j}\oint W^R_G(x,u)^{1+(1/h)}
\label{34}
\ee
and $\eta_{ij}=\delta_{e_i+e_j,h}$ can be obtained by adjustment of
$c_i$ and $b_{n,i}$.  The primary fields generate the closed operator
algebra
\be
\phi_i^R(x)\phi_j^R(x)=\sum_1^rC^k_{ij}(T)\phi^R_k(x)+Q^R_{ij}(x)
\partial_xW^R_G(x)
\label{35}
\ee
where
\be
\frac{\partial^2W^R_G(x)}{\partial T_i\partial T_j}=\partial_xQ^R_{ij}(x)
\label{36}
\ee
Note that it is better to use $X$ here instead of $x$ since we are
dealing with dispersionless or Whitham times.  Now since the special Prym
is ``universal" (cf. \cite{de,da}) the structure constants $C^k_{ij}$ are 
independent of $R$ since $C_{ijk}=C^{\ell}_{ij}\eta_{\ell k}$ is given
via $C_{ijk}(T)=<\phi_i^R\phi_j^R\phi_k^R>$.  In 2-D TFT one then has
a free energy $F$ such that $C_{ijk}=\partial^3F/\partial T_i\partial T_j
\partial T_k$ (cf. \cite{Aa,cw,dw,dc}).
\\[3mm]\indent
Now $\lambda_{SW}=[x\partial_xW/\sqrt{W^2-4\mu^2}]dx$ (for $W\sim W_G^R$) and 
one has
\be
\frac{\partial\lambda_{SW}}{\partial T_i}=-\frac{1}{\sqrt{W^2-4\mu^2}}\frac
{\partial W}{\partial T_i}dx+d\left(\frac{x}{\sqrt{W^2-4\mu^2}}
\frac{\partial W}{\partial T_i}\right)
\label{37}
\ee
Suppose $W$ is quasihomogeneous leading to
\be
x\partial_xW+\sum_1^rq_iT_i\frac{\partial W}{\partial T_i}=hW
\label{38}
\ee
($q_i=e_i+1$ is the degree of $T_i$).  Then
\be
\lambda_{SW}-\sum_1^rq_iT_i\frac{\partial\lambda_{SW}}{\partial T_i}
=\frac{hW}{\sqrt{W^2-4\mu^2}}
\label{39}
\ee
and applying $\sum q_jT_j(\partial/\partial T_j)$ to both sides yields
\be
\left(\sum_1^rq_iT_i\frac{\partial}{\partial T_i}\right)^2\lambda_{SW}-
4\mu^2h^2\frac{\partial^2\lambda_{SW}}{\partial T_r^2}=0
\label{40}
\ee
The second term represents the scaling violation due to $\mu^2=\Lambda^{2h}$
since (\ref{40}) reduces to the scaling relation for $\lambda_{SW}$ in the
classical limit $\mu^2\to 0$.  Note that $\lambda_{SW}(T_i,\mu)$ is of
degree one (equal to the mass dimension) which implies $\left(
\sum_1^rq_iT_i(\partial/\partial T_i)+h\mu(\partial/\partial\mu)-1\right)
\lambda_{SW}=0$ from which (\ref{40}) can also be obtained.  Another set of
differential equations for $\lambda_{SW}$ is obtained using (\ref{35}), to wit
\be
\frac{\partial^2}{\partial T_i\partial T_j}\lambda_{SW}=\sum_kC^k_{ij}(T)
\frac{\partial^2}{\partial T_k\partial T_r}\lambda_{SW}
\label{41}
\ee
Then the PF equations (based on (\ref{39}) and (\ref{41})) for the SW
period integrals $\Pi=\oint\lambda_{SW}$ are nothing but the Gauss-Manin
differential equations for period integrals expressed in the flat
coordinates of topological LG models.  These can be converted into $u_k$
parameters (where $\partial u_k/\partial T_r=-\delta_{kr}$) as
\be
{\cal L}_0\Pi\equiv\left(\sum_1^rq_iu_i\frac{\partial}{\partial u_i}-1
\right)^2\Pi-4\mu^2h^2\frac{\partial^2\Pi}{\partial u_r^2}=0;
\label{42}
\ee
$${\cal L}_{ij}\Pi\equiv\frac{\partial^2\Pi}{\partial u_i\partial u_j}+\sum_1^r
A_{ijk}(u)\frac{\partial^2\Pi}{\partial u_k\partial u_r}+\sum_1^r
B_{ijk}(u)\frac{\partial\Pi}{\partial u_k}=0$$
where
\be
A_{ijk}(u)=\sum_1^r\frac{\partial T_m}{\partial u_i}\frac{\partial T_n}
{\partial u_j}\frac{\partial u_k}{\partial T_{\ell}}C^{\ell}_{mn}(u);\,\,
B_{ijk}(u)=-\sum_1^r\frac{\partial^2T_n}{\partial u_i\partial u_j}\frac
{\partial u_k}{\partial T_n}
\label{43}
\ee
which are all polynomials in $u_i$.  One can emphasize that the PF equations
in 4-D $N=2$ YM are then essentially governed by the data in 2-D topological
LG models.

\section{HAMILTONIANS OF HYDRODYNAMIC TYPE}
\renewcommand{\theequation}{5.\arabic{equation}}\setcounter{equation}{0}

We have been omitting an important connection of Whitham equations and
TFT to Hamiltonians of hydrodynamic type (cf. \cite{dc,Dy}).  We think
of $F(T)$ now as a primary free energy for a TFT with $c_{ijk}=
\partial_i\partial_j\partial_kF(T)=F_{ijk}$ where $T\sim (T_n)$ 
represent Whitham or dispersionless times.  One writes $\eta_{ij}=
\eta_{ji}=c_{1ij}$ (which is assumed constant) and $c^i_{jk}=
\eta^{ip}c_{jkp}\equiv c_{ijk}=c^p_{ij}\eta_{pk}$ where $(\eta^{ij})
=(\eta_{ij})^{-1}$.  There are then WDVV equations $(\#)\,\,c^k_{ij}
c^m_{kn}=c^m_{ik}c^k_{jn};\,\,c^i_{1j}=\delta^i_j$ reflecting associativity
in the underlying TFT.  We omit references to much of the theory (cf.
\cite{cw,dc}) in order to simply exhibit 
some aspects of deformation theory, WDVV, and Hamiltonian structure for
comparison with Sections 2 and 7.  Thus we look at
systems of quasilinear PDE of hydrodynamic type ($\partial_k=\partial/\partial
T_k$)
\be
\partial_kv^p=c^p_{kq}\partial_Xv^q
\label{74}
\ee
As examples one has dKdV ($u_T=uu_X$) or
the Whitham equations from \cite{cc,fc}, namely $\partial_T\Lambda_j=
v_j(\Lambda)\partial_X\Lambda_j$ for finite zone KdV situations with branch
points $\Lambda_j$ and $T\sim T_3$.  Such equations (\ref{74}) arise 
basically from ($P=S_X$)
\be
\partial_k\lambda(X,P)=\{\lambda,\rho_k\}(X,P)
\label{75}
\ee
where $\{f,g\}=f_Pg_X-f_Xg_P$ and $L^k_{+}=B_k\to {\cal B}_k\sim\rho_k$
in a KP format
(cf. \cite{cc,ch,ci,cj,ke,ta}).  Then one has Hamiltonian equations 
(two structures) for integrable hierarchies (cf. \cite{cm,Dz}) and
averaging or taking quasiclassical limits in such equations leads to a 
compatible pair of Poisson brackets
\be
\{v^p(X),v^q(Y)\}_1=\eta^{ps}(v(X))[\delta^q_s\partial_X\delta(X-Y)-
\gamma^q_{sr}(v)v^r_X\delta(X-Y)];
\label{76}
\ee
$$\{v^p(X),v^q(Y))\}_2=g^{ps}[\delta^q_s\partial_X\delta(X-Y)-
\Gamma^q_{sr}(v)v^r_X\delta(X-Y)]$$
where the $v^p$ are arbitrary coordinates on a finite dimensional space
$M$ (which basically corresponds to a moduli space).  Here
$\eta^{pq}$ and $g^{pq}$ are contravariant components of two metrics on
$M$ and $\gamma^q_{pr}$ and $\Gamma^q_{pr}$ are the Christoffel symbols
of the corresponding Levi-Civit\`a connections.  The metric $\eta_{ab}$ is
obtained from the semiclassical limit of the first Hamiltonian structure of
an original hierarchy.  From general theory one knows that both metrics
on $M$ have zero curvature so local flat coordinates $f^1,\cdots, f^n$
on $M$ exist such that $\eta^{pq}(v)$ is constant and
$(\partial f^a/\partial v^p)(\partial f^b/\partial v^q)\eta^{pq}=
\eta_{ab}=constant$.  In these coordinates $\{f^a(X),f^b(Y)\}_1=
\eta^{ab}\delta'(X-Y)$.  To find $c^i_{jk}$ one uses the free energy
$F=log(\tau_{Whit})$ for basic times $T_k$. 
The $f^i\sim T^i$ for primary fields, to which $F$ refers (cf. 
Remark 6.1). 
Note in coupling to
topological gravity additional times arise associated to descendent fields.
The approach of \cite{dc} now is to start from a so called Frobenius
manifold $M$ and consider it as the matter sectior of a 2-D TFT; then via the
use of Hamiltonian systems of hydrodynamic type one looks for the tree
level (genus zero) approximation of a complete model obtained by coupling
the matter sector to topological gravity.  One can in fact obtain
suitable hierarchies given any solution of WDVV and the tree level
free energy is identified with the tau function of a particular solution of
the hierarchy.  We will sketch the results 
mainly for primary fields and times without going too much into the background
theory (cf. \cite{cw,dc,Dy} for more details).  Thus let $c^i_{jk}(f),\,\,
\eta_{ij}$ be a solution of WDVV where $f=(f^1,\cdots,f^n)$.  One 
constructs a solution $f^j(T),\,\,T=(T^{\alpha,p})$ where
$\alpha=1,\cdots,n$ and $p=0,1,\cdots$ of
\be
\frac{\partial}{\partial T^{\alpha,p}}f^{\beta}=c_{(\alpha,p)\gamma}^
{\beta}(f)\partial_Xf^{\gamma};\,\,T^{1,0}=X
\label{77}
\ee
The variable $X$ is usually called a cosmological constant.  Next one
determines Hamiltonian densities via
\be
\partial_{\beta}\partial_{\gamma}h_{\alpha}(f,z)=zc^{\epsilon}_
{\beta\gamma}(f)\partial_{\epsilon}h_{\alpha}(f,z);\,\,h_{\alpha}(f,0)
=f_{\alpha}=\eta_{\alpha\beta}f^{\beta};
\label{78}
\ee
$$<\nabla h_{\alpha}(f,z),\nabla h_{\beta}(f,-z)>=\eta_{\alpha\beta};\,\,
\partial_1h_{\alpha}(f,z)=zh_{\alpha}(f,z)+\eta_{1\alpha}$$
Here $\nabla$ refers to the Levi-Civit\`a connection for the invariant
metric $<\,\,\,,\,\,\,>$.  Then set $H_{\alpha,p}=\int h_{\alpha,p+1}
(f(X))dX$ and the system in (\ref{77}) has the form
\be
(\partial/\partial T^{\alpha,p})
f^{\beta}=\{f^{\beta}(X),H_{\alpha,p}\};\,\,
\{f^{\alpha}(X),f^{\beta}(Y)\}=\eta^{\alpha\beta}\delta'(X-Y)
\label{80}
\ee
Further the Hamiltonians commute pairwise and the functionals
$H_{\alpha,-1}=\int f_{\alpha}(X)dX$ span the annihilator of
the Poisson bracket in (\ref{80}).  There is a scaling group
$T^{\alpha,p}\to cT^{\alpha,p}$ with $f\to f$ for (\ref{80}) and one takes
for $f$ the invariant solution for the symmetry $[(\partial/\partial T^{1,1})-
\sum T^{\alpha,p}\partial/\partial T^{\alpha,p})]f(T)=0$.  This can be
found from the fixed point equation
\be
f=\nabla\Phi_T(f);\,\,\Phi_T(f)=\sum_{\alpha,p} T^{\alpha,p}
h_{\alpha,p}(f)
\label{81}
\ee
One now defines $<\phi_{\alpha,p}\phi_{\beta,q}>(f)>$ via
\be
(z+w)^{-1}[<\nabla h_{\alpha}(f,z),\nabla h_{\beta}(f,w)>-\eta_
{\alpha\beta}]=
\label{82}
\ee
$$=\sum_{p,q=0}^{\infty}<\phi_{\alpha,p}\phi_{\beta,q}>(f)z^pw^q=
<\phi_{\alpha}(z)\phi_{\beta}(w)>(f)$$
The infinite matrix $(<\phi_{\alpha,p}\phi_{\beta,q}>)$ represents the
EM tensor of the commutative Hamilton hierarchy (\ref{80}).  This means
that $<\phi_{\alpha,p}\phi_{\beta,q}>$ is the density of flux of 
$H_{\alpha,p}$ along the flow $T^{\beta,q}$, i.e. $(\partial/
\partial T^{\beta,q})h_{\alpha,p+1}=\partial_X<\phi_{\alpha,p}
\phi_{\beta,q}>(f)$.  Then one defines
\be
log(\tau(T))=\frac{1}{2}\sum <\phi_{\alpha,p}\phi_{\beta,q}>(f(T))
T^{\alpha,p}T^{\beta,q}+
\label{83}
\ee
$$+\sum <\phi_{\alpha,p}\phi_{1,1}(f(T))T^{\alpha,p}+\frac{1}{2}
<\phi_{1,1}\phi_{1,1}>(f(T))$$
It follows that 
\be
\frac{\partial}{\partial T^{\alpha,p}}\frac{\partial}{\partial T^{\beta,q}}
log(\tau)=<\phi_{\alpha,p}\phi_{\beta,q}>
\label{84}
\ee
Finally let ${\cal F}(T)=log(\tau(T))$ with 
$<\phi_{\alpha,p}\phi_{\beta,q}\cdots>_0
=(\partial/\partial T^{\alpha,p})(\partial/\partial T^{\beta,q})\cdots
{\cal F}(T)$.  Then 
\be
\left.{\cal F}(T)\right|_{T^{\alpha,p}=0\,(p>0);\,T^{\alpha,0}=f^{\alpha}}
=F(f)
\label{85}
\ee
along with other equations (cf. \cite{dc}).  In any event one obtains
a solution to WDVV defining coupling to topological gravity at tree level.
Note that the flat coordinates $f^{\alpha}$ are exactly the $T^{\alpha,0}$
describing primary fields and could be denoted by $T^{\alpha}$ (small
phase space).  The notion $t\sim f$ in \cite{dc,Dy} has always seemed
confusing since $t$ is used for times in the associated integrable
hierarchy such as nKdV (whereas $T\sim$ dispersionless times for dnKdV).
\\[3mm]\indent {\bf REMARK 5.1.}$\,\,$  It is clear now that we must take
another look at the Whitham times $T_n\sim T^n\sim f^n$ and $T^{\alpha,p}$.
In this direction consider the $A_n$ LG model where
\be
M=\{w(P)=P^{n+1}+g_1P^{n-1}+\cdots + g_n\}
\label{86}
\ee
Here the $g_i\in {\bf C}$ are the deformation parameters or coupling 
constants and they correspond to the $u_i$ of (\ref{30}).
Without touching axioms or definitions here one knows that $M$ is to be
a Frobenius manifold (FM) with Frobenius algebra (FA) $A=A_w$ given
by
\be
A_w={\bf C}[P]/\{w'(P)=0\};\,\,<f,g>=Res_{\infty}\frac{f(P)g(P)}{w'(P)}
\label{87}
\ee
Assuming simple distinct roots for $w'(P)$ one sets $u^i=w(P_i)$ where
$w'(P_i)=0\,\,(i=1,\cdots,n)$.  These provide canonical coordinates
$u^i$ for a diagonal metric $ds^2=\sum_1^n\eta_{ii}(du^i)^2$ where
$\eta_{ii}(u)=[w''(P_i)]^{-1}$.  This is in fact a flat Egoroff metric on 
$M$ (cf. \cite{cw,dc}).  
The corresponding flat coordinates on $M$ have the form
\be
f^{\alpha}=-\frac{n+1}{n-\alpha +1}Res_{\infty}w^{(n-\alpha +1)/(n+1)}(P)dP
\label{88}
\ee
where $\alpha=1,\cdots, n$ and in these coordinates $ds^2=\eta_{\alpha\beta}
df^{\alpha}df^{\beta}$ with $\eta_{\alpha\beta}=\delta_{n+1,\alpha+\beta}$.
These are called $T_{\alpha}$ in (\ref{31}) (which is standard) and
$t^{\alpha}$ in \cite{dc} (we will clarify this below).  One is dealing
here with the dispersionless limit of a KdV hierarchy based on $L=
\partial^{n+1}+g_1\partial^{n-1}+\cdots + g_n$ where $\partial=\partial/
\partial x$ and there is the background hierarchy
\be
\frac{\partial L}{\partial \tau^{\alpha,p}}=c_{\alpha,p}\left[L,
(L^{(\alpha/(n+1))+p})_{+}\right]
\label{89}
\ee
where $\alpha=1,\cdots, n$, $p=0,1,\cdots$, and the $c_{\alpha,p}$ are 
certain constants.  In the dispersionless limit one has $x\to \epsilon x
=X$ and $\tau^{\alpha,p}\to\epsilon\tau^{\alpha,p}=T^{\alpha,p}$.  The
differential equation in (\ref{78}) has a solution
\be
h_{\alpha}(t,z)=-\frac{n+1}{\alpha}Res_{\infty}\,w^{\alpha/(n+1)}{}_1
F_1\left(1,1+\frac{\alpha}{n+1},zw(P)\right)dP
\label{90}
\ee
In particular we note that the $t^{\alpha}=T^{\alpha,0}$ provide 
coordinates for $M$ just as the $g_i$ would, so both
the $g_i$ and $t^{\alpha}$ are coupling constants or moduli; 
they should have equal ``status"
in some sense.
Now to straighten out further any possible $t,T$ confusion we note
note that $c^i_{jk}(t)$ arises from considering deformations of a TFT which
preserve topological invariance (i.e. produce other TFT).  The idea here is
to capture more information about a topological Lagrangian (and about
the whole shebang) by studying topological deformations.  This leads to
the WDVV equations in the $t^{\alpha}$ variables.  Then for any solution
of WDVV one constructs a hierarchy of integrable Hamiltonian equations
of hydrodynamic type such that the tau function of a particular solution
coincides with the genus zero approximation of the corresponding TFT
model coupled to gravity.  One must now regard the flat $t^{\alpha}\sim
T^{\alpha,0}$ arising via (\ref{88}) as $f^{\alpha}$ with $c^i_{jk}(t)$
given and for $p=0,\,\,\partial f^{\beta}/\partial T^{\alpha,0}=
\delta_{\alpha\beta}$ and $\partial_Xf^{\gamma}=\delta_{\gamma 1}$ so
that (\ref{77}) becomes $\delta_{\alpha\beta}=c^{\beta}_{\alpha\gamma}
\delta_{\gamma 1}$ which implies $\delta_{\alpha\beta}=c^{\beta}_{\alpha 1}=
\eta^{\beta p}c_{\alpha 1p}=\eta^{\beta p}\eta_{p\alpha}$ which is correct.
Thus the hydrodynamic equations (\ref{77}) are nontrivial only for the
$T^{\alpha,p}$ with $p>0$.

\section{RELATIONS TO C THEOREM AND WDVV}
\renewcommand{\theequation}{6.\arabic{equation}}\setcounter{equation}{0}

\subsection{C Theorem}

We refer to \cite{By,ba,Bz,dy,do,gh,lf,mu,zc,zb} and will concentrate
on \cite{By,Bz}.  These matters seem to have been first broached 
analytically in \cite{ba} and reached a point of fruition in \cite{By,Bz}
(cf. also Section 3 based on \cite{dz,dg}).  We go first to \cite{Bz} and
look at the $SU(2)$ theory where $u=\pi i(F-(a/2)F_a)$ (cf. Section 1
with $b_1=1/4\pi^2$).  We recall also the formula for beta functions in
(\ref{D}).  Now based on \cite{ba,ma} one writes
\be
\frac{u}{\Lambda^2}=J(\tau)=2\frac{\theta_3^4}{\theta_2^4}-1
\label{102}
\ee
where $J:\,\,{\bf H}\to {\bf C}/\{\pm 1\}$ is the uniformizing map and 
${\bf H}$ is the Poincar\'e upper half plane (cf. \cite{bc,ba,ma} for the 
elliptic $\theta_i$).  Then
\be
\beta(\tau)=\Lambda\left.\frac{d\tau}{d\Lambda}\right|_u=-2\frac{J(\tau)}
{J'(\tau)}=-\frac{i}{\pi}\left(\frac{1}{\theta_3^4}+\frac{1}{\theta_4^4}
\right)
\label{103}
\ee
(cf. also \cite{do,lf,rd}) and 
\be
\frac{d\tau}{\beta(\tau)}=-\frac{J'd\tau}{2J}=-\frac{1}{2}\partial_{\tau}
log|J|d\tau=-\frac{1}{4}\partial_{\tau}log|J|^2d\tau
\label{104}
\ee
But $\Lambda\partial_{\Lambda}|J|^2=-4|J|^2$ so $|u/\Lambda^2|^2$ is
nonincreasing along the RG flow.  This means that $L_2=|J|^2=exp
(-4\Psi_2)$ is a Lyapunov function for the RG flow.
\\[3mm]\indent {\bf REMARK 6.1.}$\,\,$
In the paper \cite{lf}
on RG potentials in YM theories, one starts from
the same beta function (\ref{103}) written as 
$\beta^z=[(1-4f(z))/f'(z)]$ for $f=-(\theta_3\theta_4/
\theta_2^2)^4(z)$. Then the covariant beta function $\beta_z$
has the form
\be
\beta_z=\frac{f'}{1-4f}=\partial_z\Phi;\,\,\Phi=-\frac{1}{4}log|1-4f|^2
\label{62}
\ee
(cf. (\ref{104}))
from which one extracts a metric
\be
G_{z\bar{z}}=\beta_z\beta_{\bar{z}}=\left|\frac{f'}{1-4f}\right|^2=
\partial_z\partial_{\bar{z}}K;\,\,K=c\Phi+\frac{1}{2}\Phi^2
\label{63}
\ee
This shows that the RG flow is gradient in the sense that $-\beta^i\partial_i
\Phi=-\beta^i\beta_i=-\beta^i\beta^jG_{ij}\leq 0$ and 
it is mentioned that the
result is logically
independent of any relations to the Zamolodchikov C function.
\\[3mm]\indent
This kind of result is now extended in \cite{Bz} to $SU(3)$ and eventually
to $SU(n)$ using beta functions
$\beta_{ij}=\Lambda(\partial\tau_{ij}/\partial\Lambda)|_u$ where
$\tau_{ij}=\oint_{B_j}d\omega_i=\partial^2F/\partial a_i\partial a_j$.
One writes
\be
\Delta^{SU(n)}_{cl}(u^{\gamma})=\prod_{i<j}^n(e_i-e_j)^2
\label{105}
\ee
where the $e_i$ are the zeros in $x$ of the polynomial $W_{A_{n-1}}(x,u^2,
\cdots,u^n)=x^n-\sum_2^nu^{\gamma}x^{n-\gamma}$ (cf. (\ref{30})).  One
defines $b=\beta^{ij}d\tau_{ij}$ and $\hat{u}_{\gamma}=u^{\gamma}/
\Lambda^{\gamma}$ for $\gamma=2,\cdots,n$.  Then setting
$b=(\sum_2^nd\hat{u}_{\gamma}\partial_{\gamma})\Psi_n$ with
\be
\Psi_n=-\frac{1}{n}log\left|\frac{\Delta^{SU(n)}_{cl}(u^{\gamma})}
{\Lambda^{n(n-1)}}\right|^2=-\frac{1}{n}log\left|\hat{\Delta}^{SU(n)}_{cl}
(\tau)\right|^2
\label{106}
\ee
one has for $L_n=exp(-n\Psi_n)$
\be
L_n=e^{-\Psi_n}=\left|\hat{\Delta}_{cl}^{SU(n)}(\tau)\right|^2;\,\,
\Lambda\partial_{\Lambda}L_n=-n(n-1)L_n
\label{107}
\ee
The meaning here is that the classical symmetry restoring locus plays a
nontrivial attracting role in the theory.

\subsection{WDVV}

We go next to \cite{By} and refer to \cite{ba,cw,dw,dc,Dy,ko,kz,My,Mw,Mz}
for WDVV.  The reult in \cite{By} is that $\beta^{ij}=\eta^{ij}$ where
$\eta^{ij}=(\eta_{ij})^{-1}$ corresponds to the WDVV metric and 
$\beta^{ij}=(\beta_{ij})^{-1}$.  An offshoot is the natural
conjecture that $u=(i/4\pi b_1)(F-\sum (a_i/2)a_i^D)$ is equivalent to
WDVV in the form
\be
F_{ik\ell}\beta^{\ell m}F_{mnj}=F_{jk\ell}\beta^{\ell m}F_{mni}
\label{108}
\ee
Recall here in Section 5 we wrote (for a one puncture situation)
\be
c^i_{jk}=\eta^{ip}c_{jkp};\,\,c_{ijk}=c^p_{ij}\eta_{pk};\,\,
c_{ijk}=\partial_i\partial_j\partial_kF;\,\,\eta_{ij}=c_{1ij}
\label{109}
\ee
so $\beta^{ij}=\eta^{ij}$ suggests
\be
\beta_{ij}=\Lambda\partial_{\Lambda}\left.\tau_{ij}\right|_u=\Lambda
F_{\Lambda ij}=c_{1ij}=F_{1ij}
\label{110}
\ee
provided one can isolate the ``puncture operator" corresponding to 
$\partial_1$.  Note that this may perhaps not be the $\partial_1$ (or 
$\partial_r$ in the notation of (\ref{30})) which is standard in TFT
or LG models. 
Note also that in \cite{By} one does not define the $\eta_{ij}$ or
$\beta_{ij}$ via differential forms and only the $a_j$ variables are 
involved (not the $T_n$).  Thus the relation of $F$ here to the
$F$ of \cite{cc,ia,ka,ko,na} or to that of \cite{My,Mw,Ms,Mz} is not clear.
The matter will be partially clarified in what follows
(cf. also Section 3).  Going to \cite{ko}
one takes a RS $\Sigma_g$ of genus $g$ with $N$ punctures $P_{\alpha}$.
Pick Abelian differentials $dE$ and $dQ$ such that $E$ and $Q$ have poles
of order $n_{\alpha}$ and $m_{\alpha}$ respectively at $P_{\alpha}$ and set
$d\lambda=QdE$ with a pole of order $n_{\alpha}+m_{\alpha}+1$ at $P_{\alpha}$
(this corresponds to the SW differential).  Pick local coordinates $z_{\alpha}$
near $P_{\alpha}$ so that $E\sim z_{\alpha}^{-n_{\alpha}}+R^E_{\alpha}log
(z_{\alpha})$, require $\oint_{A_j}dQ=0$, and fix the additive constant
in $\lambda$ by requiring that its expansion near $P_1$ have no constant
term.  Define times
\be
T_{\alpha,i}=-\frac{1}{i}Res_{P_{\alpha}}(z_{\alpha}^id\lambda)\,\,
(1\leq \alpha\leq N,\,\,1\leq i\leq n_{\alpha}+m_{\alpha});
\,\,R_{\alpha}^{\lambda}=Res_{P_{\alpha}}(d\lambda)
\label{WJ}
\ee
where $2\leq \alpha\leq N$ in the last set.
This gives $\sum_1^N(n_{\alpha}+m_{\alpha})+N-1$ parameters.  The remaining
parameters needed to parametrize 
the space ${\cal M}_g(n,m)$ of the creatures indicated consist of the
$2N-2$ residues of $dE$ and $dQ$, namely $R^E_{\alpha}=Res_{P_{\alpha}}dE$
and $R^Q_{\alpha}=Res_{P_{\alpha}}dQ\,\,(2\leq \alpha\leq N)$, plus
$5g$ parameters
\be
\tau_{A_i,E}=\oint_{A_i}dE;\,\,\tau_{B_i,E}=\oint_{B_i}dE;\,\,
\tau_{A_i,Q}=\oint_{A_i}dQ;\,\,\tau_{B_i,Q}=\oint_{B_i}
dQ;\,\,a_i=\oint_{A_i}QdE
\label{WK}
\ee
where $1\leq i\leq g$ in the last set.  Then it is proved in 
\cite{ka} that, if ${\cal D}$ is the open set in ${\cal M}_g(n,m)$ where
the zero divisors $\{z;\,dE(z)=0\}$ and $\{z;\,dQ(z)=0\}$ do not intersect,
then the joint level sets of the set of all parameters except the $a_i$
define a smooth $g$-dimensional foliation of ${\cal D}$.  Further near
each point in ${\cal D}$ the $5g-3+3N+\sum_1^N(n_{\alpha}+m_{\alpha})$
parameters $R^E_{\alpha},\,R^Q_{\alpha},\,F^{\lambda}_{\alpha},\,T_{\alpha,k},\,
\tau_{A_i,E},\,\tau_{B_i,E},\,\tau_{A_i,Q},\,\tau_{B_i,Q},$ and $a_i$
have linearly independent differentials and thus define a local holomorphic
coordinate system.
Assume now that $dE$ has simple zeros $q_s\,\,(s=1,\cdots,2g+n-1$ in
the case of interest ${\cal M}_g(n,1)$ below
since for one puncture $\#(zeros)-n-1=2g-2$
by Riemann-Roch) and we come to the Whitham
times.
The idea here is that suitable submanifolds of ${\cal M}_g(n,m)$ are
parametrized by $2g+N-1+\sum_1^N(n_{\alpha}+n_{\alpha})$ Whitham
times $T_A$ to each of which is associated a dual time $T^D_A$ and
an Abelian differential $d\Omega_A$.  First take $N=1$ (one puncture)
with 
\be
T_j=-\frac{1}{j}Res(z^jd\lambda);\,\,T^D_j=Res(z^{-j}d\lambda);\,\,d\Omega_j
\,\,(1\leq j\leq n+m)
\label{WS}
\ee
($d\Omega_j=d(z^{-j}+O(z))$ with $\oint_{A_j}d\Omega_i=0$).
For $g>0$ there are $5g$ more parameters and we consider only foliations
for which $\oint_{A_k}dE,\,\,\oint_{B_k}dE$, and $\oint_{A_k}dQ$ are fixed.
This leads to 
\be
a_k=\oint_{A_k}d\lambda;\,\,T^E_k=\oint_{B_k}dQ;\,\,a^D_k=
-\frac{1}{2\pi i}\oint_{B_k}d\lambda;\,\,{}^DT^E_k=\frac{1}{2\pi i}
\oint_{A_k^{-}}Ed\lambda
\label{WT}
\ee
The corresponding differentials are $d\omega_k$ and $d\Omega^E_k$ where 
the $d\Omega^E_k$ are holomorphic on $\Sigma$ except along $A_j$ cycles
where $(\bullet\spadesuit)\,\,d\Omega^{E_{+}}_k-d\Omega^{E_{-}}_k=
\delta_{jk}dE$.  Thus one has $2g+n+m$ times $T_A=(T_j,a_k,T^E_k)$ 
and for $N>1$ punctures there are $2g+\sum (n_{\alpha}+m_{\alpha})$ times
$(T_{\alpha,j},a_k,T^E_k)$ plus $3N-3$ additional parameters for the residues
of $dQ,\,dE,$ and $d\lambda$ at the $P_{\alpha}\,\,(2\leq\alpha\leq N)$.
For convenience one considers only leaves where $(\bullet\diamondsuit\bullet)
\,\,Res_{P_{\alpha}}dQ=0;\,\,Res_{P_{\alpha}}dE=\,\,fixed\,\,(2\leq\alpha
\leq N)$ and incorporates among the $T_A$ the residues $R_{\alpha}^{\lambda}=
Res_{P_{\alpha}}d\lambda\,\,(2\leq\alpha\leq N)$ with $N-1$ dual times
$(\bullet\heartsuit\bullet)\,\,{}^DR^{\lambda}_{\alpha}=-\int_{P_1}^{P_{\alpha}}
=\lambda_{\alpha}$ where $2\leq\alpha\leq N$, corresponding to differentials
$d\Omega^3_{\alpha}$ which are Abelian differentials of third kind with 
simple poles at $P_1$ and $P_{\alpha}$ and residue $1$ at $P_{\alpha}$.  The
Whitham tau function is $\tau=exp({\cal F}(T))$ where
\be
{\cal F}(T)=\frac{1}{2}\sum_AT_AT^D_A+\frac{1}{4\pi i}\sum_1^ga_k
T^E_kE(A_k\cap B_k)
\label{WU}
\ee
Here $A_k\cap B_k$ is the point of intersection of these cycles.  When
$Res_{P_{\alpha}}dE=0$ one obtains the derivatives of ${\cal F}$ with
respect to the $2g+\sum(n_{\alpha}+m_{\alpha})+N-1$ Whitham times as
\be
\partial_{T_A}{\cal F}=T^D_A+\frac{1}{2\pi i}\sum_1^g\delta_{a_k,A}
T^E_kE(A_k\cap B_k);
\label{WV}
\ee
$$\partial^2_{T_{\alpha,i},T_{\beta,j}}{\cal F}
=Res_{P_{\alpha}}(z_{\alpha}^id\Omega_
{\beta,j});\,\,\partial^2_{a_j,A}{\cal F}=\frac{1}{2\pi i}\left(E(A_k\cap B_k)
\delta_{(E,k),A}-\oint_{B_k}d\Omega_A\right);$$
$$\partial^2_{(E,k),A}{\cal F}=\frac{1}{2\pi i}\oint_{A_k}Ed\Omega_A;\,\,
\partial^3_{ABC}{\cal F}=\sum_{q_s}Res_{q_s}\left(\frac{d\Omega_A
d\Omega_Bd\Omega_C}{dEdQ}\right)$$
When $Res_{P_{\alpha}}dE\not= 0$ some modifications are needed.
For one puncture the case of interest here is $Q_{+}=z^{-1}$ and there
are two Whitham times $T_n=0$ and $T_{n+1}=n/(n+1)$ fixed so we will
have $2g+n-1$ Whitham times for ${\cal M}_g(n,1)$.
Next one shows that each $2g+n-1$ dimensional leaf $\hat{{\cal M}}$ of the 
foliation of ${\cal M}_g(n)$ parametrizes the marginal deformation of a
TFT on $\Sigma$.  The free energy of such theories is the restriction
to the leaf of ${\cal F}$. Thus we consider the leaf within ${\cal M}_g(n,1)$
of dimension $2g+n-1$ which is defined by the constraints
\be
T_n=0;\,\,T_{n+1}=\frac{n}{n+1};\,\,\oint_{A_k}dE=0;\,\,\oint_{A_k}dQ=0;\,\,
\oint_{B_k}dE= fixed
\label{111}
\ee
Thus the leaf is parametrized by the $n-1$ Whitham times $T_A\,\,(A=1,
\cdots,n-1)$ and by the periods $a_k=\oint_{A_k}d\lambda$ and 
$T^E_k=\oint_{B_k}dQ$.  There will be primary fields $\phi_i\sim
d\Omega_i/dQ\,\,(i=1,\cdots,n-1)$ plus $2g$ additional fields
$d\omega_i/dQ$ and $d\Omega^E_j/dQ$. 
Then one can define
\be
\eta_{A,B}=\sum_{q_s}Res_{q_s}\frac{d\Omega_Ad\Omega_B}{dE};\,\,
c_{ABC}=\sum_{q_s}Res_{q_s}\frac{d\Omega_Ad\Omega_Bd\Omega_C}{dEdQ}
\label{112}
\ee
where $T_A\sim (T_i,a_j,T^E_k)$.  
The formulas (\ref{WV}) hold as before and the Whitham equations 
are generically 
$\partial_Ad\Omega_B=\partial_Bd\Omega_A$ which can in
fact be deduced from $\partial_AE=\{\Omega_A,E\}$ where $\{f,g\}=f_pg_X
-g_pf_X$ with $dp\sim d\Omega_1$ (cf. \cite{cc,ka,kc,ko}).
We see that for $A=1,\,\,d\Omega_A=dQ$ implies formally $c_{1BC}=\eta_{BC}$
so $T_1$ plays a special role in the general theory
and one can imagine heuristically that the
role of $\Lambda\partial_{\Lambda}$ 
in (\ref{110}) is formally the same as 
$\partial_1$ when acting on $F$.  This suggests
$(\bullet\clubsuit\bullet)\,\,T_1\sim log(\Lambda)$ but this is
really only a one puncture argument ($P_{+}\sim\infty$),
and moreover in the present context $\eta_{ij}=0$ for $i,j\sim a_i,a_j$ (see
below).
\\[3mm]\indent {\bf REMARK 6.2.}$\,\,$  In Remark 1.2 we have 
$\Lambda\partial_{\Lambda}\sim T_1\partial_1$ which holds when, as in
Remark 1.1, $T_1\sim c\Lambda$.  However it is easier to regard the
situation of Remarks 1.1 and 1.2 as expressing the effect of $T_1$ in
restoring homogeneity to the prepotential
and Remark 3.1 suggests that more $T_j$ are generally needed.  
The identification $T_1\sim 
log(\Lambda)$ of $(\bullet\clubsuit\bullet)$
would be more direct and substantial 
in distinguishing a special role for $T_1=X$ but the
case of one puncture is artificial in the SW theory and
we should better use two punctures $P_{\pm}\sim\infty_{\pm}$ in the
Toda context 
which means $X$ does not correspond to $dQ$ (so $X$ is no longer
distinguished).
We note also that in
the second paper of \cite{By} the authors extend the WDVV type equations
for $N=2$ susy YM by introducing directly a new variable $a_0\sim\Lambda$
and extending the index set from $(1,n-1)$ to $(0,n-1)$ (the $T_n$
variables are completely ignored).  This is regarded as a necessary
step in looking for fully topological WDVV equations for $N=2$ SYM
(without embedding in the Whitham hierarchy).  The relation of WDVV theories
based on the $a_j$ alone (as in \cite{By,My,Mw}) to WDVV for a full
Whitham theory as in \cite{cc,dc,kc,ko} is not entirely clear.
The truncated prepotential depending on the $a_j$ alone satisfies
a wider set of WDVV equations $F_iF_k^{-1}F_j=F_jF_k^{-1}F_i$ where
$(F_i)_{mn}=F_{imn}$.  The $\eta_{ij}=F_{1ij}$ in (\ref{110}) refer to $a_i$ 
and $a_j$ but in the full (one puncture)
theory with $T_n$ etc. such an $\eta_{ij}$ 
will vanish (cf. \cite{ko}).
(note some clarification in Remark 6.3 below
for the two puncture situation).  In any case
it does not seem to be correct to extend formulas from 
the structure of \cite{By} to the full
Whitham hierarchy and $(\bullet\clubsuit\bullet)$
is unlikely.
\\[3mm]\indent {\bf REMARK 6.3.}$\,\,$  Let us try to clarify the problems
indicated in Remark 6.2 concerning $(\bullet\clubsuit
\bullet)$.  At issue here is the compatibility of WDVV built from the
$a_i$ variables alone and WDVV for the full Whitham hierarchy.  For the
full Whitham hierarchy we have indicated the construction in this section
(cf. also \cite{cc,cg,dc,ka,kc,ko}).  For the truncated but more general
WDVV theory we sketch a few points here following \cite{ba,By,My,Mw,Ms,Mz}.
Thus go to \cite{My} and look at a simple situation $(\bullet\spadesuit
\bullet)\,\,w+(1/w)=2P(\lambda),\,\,P(\lambda)=\lambda^N+\sum_1^{N-1}
s_k\lambda^{k-1},\,\,dS=\lambda(dw/w),\,\,y=(1/2)(w-(1/w)),\,\,g=N-1$
and recall from \cite{cc,ko,na} that $w=y+P$ with 
\be
dS=\frac{\lambda dP}{y}=\frac{\lambda dy}{P}=\frac{\lambda dw}{w}
\label{120}
\ee
($w\sim h$ in \cite{na}).  For TFT with a LG potential $W(\lambda)$ one writes
$\phi_i\phi_j=c^k_{ij}\phi_k\,\,mod\,W'$ with $F_{ijk}=Res[\phi_i\phi_j
\phi_k/W']=\sum[(\phi_i\phi_j\phi_k)(\lambda_{\alpha})/W''(\lambda_{\alpha})]$ 
where $W'(\lambda_{\alpha})=0$ (simple zeros).  Then $\eta_{ij}=Res
[\phi_i\phi_j/W']$ and $F_{ijk}=\eta_{k\ell}c^{\ell}_{ij}$ with $\phi_1\sim
1$.  This corresponds to standard Whitham theory type WDVV using just the 
$T_n$ times, and can be phrased via differentials $d\Omega_A$ as in Section
6.2.  For the truncated WDVV with only variables $a_i$ involved one writes
$a_i=\oint_{A_i}dS$ with $a^D_i=\oint_{B_i}dS$ and $a_i\sim d\omega_i$ where
the $d\omega_i$ are holomorphic differentials with $\oint_{A_i}d\omega_j=
\delta_{ij}$.  In the present situation one can write the $d\omega_i$ as 
linear combinations of holomorphic differentials
\be
dv_k=\frac{\lambda^{k-1}d\lambda}{y}\,\,(k=1,\cdots,g);\,\,y^2=P^2-1=\prod_1^
{2g+2}(\lambda-\lambda_{\alpha})
\label{121}
\ee
(where $g=N-1$).  Note also from (\ref{121}) that
$
2ydy=\sum_1^{2N}\prod_{\alpha\not=\beta}(\lambda-\lambda_{\alpha})d\lambda$
so $d\lambda=0$ corresponds to $y=0$.  Then in \cite{My} one defines
(recall $dw/w=dP/y$)
\be
F_{ijk}=\frac{\partial^3F}{\partial a_i\partial a_j\partial_k}=
Res_{d\lambda=0}
\left(\frac{d\omega_id\omega_jd\omega_k}{d\lambda(dw/w)}\right)=
\label{123}
\ee
$$=\sum_1^{2g+2}\frac{\hat{\omega}_i(\lambda_{\beta})\hat{\omega}_j
(\lambda_{\beta})\hat{\omega}_k(\lambda_{\beta})}{P'(\lambda_{\beta})/
\hat{y}(\lambda_{\beta})}$$
where $d\omega_i(\lambda)=[\hat{\omega}_i(\lambda_{\beta})+O(\lambda-
\lambda_{\beta})]d\lambda$ and $\hat{y}^2(\lambda_{\beta})=\prod_
{\beta\not=\alpha}(\lambda_{\beta}-\lambda_{\alpha})$.  For the metric
one takes
\be
\eta_{ij}(d\omega)=Res_{d\lambda=0}\left(\frac{d\omega_i\omega_jd\omega)}
{d\lambda (dw/w)}\right)=\sum\frac{\hat{\omega}_i(\lambda_{\beta})
\hat{\omega}_j(\lambda_{\beta})\hat{\omega}(\lambda_{\beta})}
{P'(\lambda_{\beta})/\hat{y}(\lambda_{\beta})}
\label{124}
\ee
Then the $c^k_{ij}(d\omega)$ can be obtained via
\be
F_{ijk}=\eta_{k\ell}(d\omega)c^{\ell}_{ij}(d\omega)
\label{125}
\ee
(see below).
These formulas could also be expressed via
\be
F_{ijk}=-Res_{d\,log(w)=0}\left(\frac{d\omega_id\omega_jd\omega_k}
{d\lambda (dw/w)}\right)
\label{126}
\ee
and via $dw/w=dP/y$ the calculation can be taken over
the $q_s$ where $dE(q_s)=0$.
Thus the formula of (\ref{123}) is compatible with $F_{ABC}$ of (\ref{112})
but the $\eta$ terms (\ref{112}) and (\ref{124}) are incompatible since
$\eta_{a_ia_j}=0$ in (\ref{112}).
\\[3mm]\indent
We have neglected to spell out the puncture picture completely and for
this we refer to \cite{na}; namely there are differentials $d\Omega^{+}$ and
$d\Omega^{-}$ of second kind for $i\geq 1$ associated respectively to
times $T_n$ and $\hat{T}_n$ and $d\Omega_0$ of third kind for $n=0$
associated with $T_0$.  Here one writes, based on \cite{na}, near $P_{+}$
\be
d\Omega_n^{+}=\left[-nz^{-n-1}-\sum_1^{\infty}q_{mn}z^{m-1}\right]dz\,\,
(n\geq 1);
\label{130}
\ee
$$d\Omega^{-}_n=\left[\delta_{n0}z^{-1}-\sum_1^{\infty}r_{mn}z^{m-1}\right]
dz\,\,(n\geq 0)$$
while near $P_{-}$
\be
d\Omega^{+}_n=\left[-\delta_{n0}z^{-1}-\sum_1^{\infty}\hat{r}_{mn}z^{m-1}
\right]dz\,\,(n\geq 0);
\label{131}
\ee
$$d\Omega^{-}_n=\left[-nz^{-n-1}-\sum_1^{\infty}\hat{q}_{mn}z^{m-1}\right]
dz\,\,(n\geq 1)$$
Finally $d\Omega_0$ has simple poles at $P_{\pm}$ with residues $\pm 1$
and is holomorphic elsewhere; further $d\Omega^{-}_0=d\Omega^{+}_0=
d\Omega_0$ is stipulated.  
However an attempt to define $\eta_{a_ia_j}$ by $\partial^3F/\partial T_0
\partial a_i\partial a_j$ is not successful.
Let us also establish a uniform notation connecting e.g. \cite{ka,ko} with
\cite{na}.  Thus we have been using $dS=\lambda dP/y=\lambda dw/w=
\lambda dh/h$ where $h$ in \cite{na} corresponds to $w$ in \cite{ka} 
for example so let's stay with that.  Then in \cite{na} one 
writes $y^2=P^2-\Lambda^{2N},\,\,P=\lambda^N+\sum_0^{N-2}u_{N-k}\lambda^k,\,\,
h=y+P,\,\,\tilde{h}=-y+P,$ and $h\tilde{h}=\Lambda^{2N}$ with 
$h^{-1}\sim z^N$ at $P_{+}$ and $z^N=\tilde{h}^{-1}$ at $P_{-}$ for a local
coordinate $z$ (actually $Q\sim\lambda\sim (1/z)$ here as indicated below).
Further $div(h)=NP_{-}-NP_{+},\,\,div(\tilde{h})=NP_{+}-NP_{-},$ and 
$Res_{+}h^{n+1}QdE+Re_{-}h^{n+1}QdE=0$ for all $n$.  In \cite{ko} one takes 
$dE\sim dP/y$ with two simple poles at $P_{\pm}$ having residues $-N$ and $N$
respectively while $Q$ is a meromorphic function with poles at $P_{\pm}$
which plays the role of coordinate $Q\sim 1/z$ in each sheet (more below).
Further one writes $y^2=\prod_1^N(Q-\bar{a}_k)^2-\Lambda^{2N}=P^2
-\Lambda^{2N}$ (so $Q\sim \lambda$) and $E=
log(y+P)$ which means $E\sim log(h)=log(w)$ with $dE=dh/h=dw/w$ as
above.  Note also e.g. $Z^N\sim h^{-1}$ at $P_{+}$ corresponds to $Nlog(z)
=-log(h)\sim -E$ or $E=log(h)$.  Finally in \cite{ka} one takes 
$w=h=exp(E),\,\,w+(\Lambda^{2N}/w)=2P(Q),$ and $2y=w-(\Lambda^{2N}/w)$ so
$y^2=P^2-\Lambda^{2N}$ and $h\sim w\in {\bf P^1}$ is the Toda variable
(cf. also \cite{ia}).  This is related to $\lambda$ arising in an equation
$det(\lambda-L(h))=0$ for example.  Further near $P_{+}$
\be
E\sim -Nlog(z);\,\,Q\sim 2^{-1/N}z^{-1}+O(1)
\label{133}
\ee
so $E=Nlog(Q)+log(2)+O(Q^{-1})$,
while near $P_{-},\,\,Q\sim (\Lambda/2)^{1/N}z^{-1} +O(1)$ with
\be
E=-Nlog(Q)+log(2)+log\left(\frac{\Lambda}{4}\right)+O(Q^{-1})
\label{134}
\ee
\\[3mm]\indent
It seems now that linking of $F_{ijk}=\eta_{k\ell}(d\omega)c^{\ell}_{ij}
(d\omega)$ as in (\ref{123}) or (\ref{126}) and $F_{ABC}=c_{ABC}$ as in
(\ref{112}) must perhaps be undertaken in the context of a deformation theory.
Perhaps something like $F_{ABC}^{ijk}=
F_{ijk}(a_m)+\tilde{F}_{ABC}(a_m,T_n)$ would
do with $\Lambda$ arising either in $F_{ijk}$ as in \cite{By} or via
$\tilde{F}_{ABC}$ somehow.
In terms of special geometry (cf. \cite{ca,fg,Ga,kz,la}) one is tempted
to think of the $a_i$ variables in terms of K\"ahler deformations
and the $T_n$ in terms of deformations of complex structure.
In \cite{By} one introduces a new variable $a_0$ which plays the role of
$\Lambda$ where $\beta_{ij}=\Lambda\partial_{\Lambda}\tau_{ij}=
\Lambda\partial_{\Lambda}(\partial^2F/\partial a_i\partial a_j)$
with $i,j=1,\cdots,g$ and $\partial_0F=\partial_{\Lambda}F$.  In effect
what this does is allow one to restore the homogeneity of $F$ in the form
$\sum_0^ga_j(\partial F/\partial a_j)=2F$ without using the Whitham
times.  In a sense this may make the use of Whitham deformations unnecessary
but it does not deal with the natural role of 
Whitham times connected to SW theory.
There seems then to be two different formulations of WDVV; one, $(WDVV)_{Whit}$,
for a full SW-Whitham theory, which upon restriction to pure SW theory does
not reduce to the second, $(WDVV)_{SW}$, defined only on the variables
$a_i\,\,(0\leq i\leq g)$ as in \cite{By} or (\ref{123}) - (\ref{124}). 
The homogeneity of $F$, which is disturbed
by renormalization, can be controlled in $(WDVV)_{SW}$ via $a_0\sim
\Lambda$, or in $(WDVV)_{Whit}$ by use of $T_n$ variables 
(with $a_0$ absent), possibly via some variation on $(\bullet\clubsuit
\bullet)$.  Arguments in special geometry (cf. \cite{fg}) suggest that
renormalization changes the K\"ahler metric but a superpotential 
remains unrenormalized, so one is let to question Whitham times as
renormalization parameters (cf. Section 8). 
Let us write
e.g. $F_{ijk}\sim F_{SW}$ as in (\ref{123}) and $\eta_{ij}(d\omega)$ as in
(\ref{124}) with $\eta_{AB}$ and $F_{ABC}\sim F_{Whit}$ as in (\ref{112})
(but now based on a two puncture situation $P_{\pm}\sim\infty_{\pm}$ as
in \cite{ko,na}).  
One might consider $F=F_{SW}(a) +F_{Whit}(\hat{a},T)$ for
$a\sim (a_0=\Lambda,a_1,\cdots,a_g)$ and $\hat{a}\sim (a_1,\cdots,a_g)$  
and ask whether this leads to WDVV 
for $F$ in some suitable sense.   Note from \cite{ko}
for $F_{Whit},\,\,\eta_{ij}=\delta_{i+j,0}$ for $i,j\sim d\Omega_i^{\pm},\,
d\Omega_j^{\pm}$ with $\eta_{a_i,(E,k)}=\delta_{ik}$ where 
$T_k^E$ is defined in (\ref{WT}); all other pairings vanish so in the metric the
linking of the $a_j$ to general $T_A$ in $F_{Whit}$ occurs only via $T_k^E$.  
In any event expressions via residues in the two puncture 
situation lead to satisfactory third derivatives.
We note that $\partial^2F/\partial a_i\partial a_j$ is not
calculated via formulas such as (\ref{112}) so no information is thence
available.

\section{DEFORMATIONS}
\renewcommand{\theequation}{7.\arabic{equation}}\setcounter{equation}{0}

We extract here from \cite{lg}.  The subject is topological
gauge theories (TGT), arising from general $N=2$ twisted gauge theories,
studied in the Gromov-Witten (GW) paradigm.  We will not discuss
the GW ideas or Donalson theory but only look at some properties of
the prepotential ${\cal F}$ which 
form a small part of the technique in \cite{lg}. We
look at the standard $SU(2)$ moduli space ${\cal M}$ with $u\in{\cal M}$
and ${\cal F}={\cal F}(a,t_r)$ is called a master function ($\sim$
prepotential); here $a(u)$ and $a^D(u)$ are the standard ``moduli parameters"
satisfying
\be
\left[(1-u^2)\frac{d^2}{du^2}+\frac{1}{4}\right]\left(
\begin{array}{c}
a(u)\\
a^D(u)
\end{array}\right)=0
\label{44}
\ee
with asymptotics as $u\to\infty$ given by 
$a(u)\sim\sqrt{u/2}+\cdots$ and $a^D\sim-\frac{2a}{\pi i}log(u)+\cdots$.
Then
${\cal F}(a,0)$ is defined as the solution of $d{\cal F}=a^D(u)da(u)$ and
a function $H(a,a^D)$ is defined via
\be
H(a(u),a^D(u))=-\frac{u}{2};\,\,H(\mu a,\mu a^D)=\mu^2H(a,a^D)\,\,(\mu\not= 0)
\label{46}
\ee
$$H(\gamma a^D+\delta a,\alpha a^D+\beta a)=H(a,a^D);\,\,
\left(\begin{array}{cc}
\alpha & \beta\\
\gamma & \delta
\end{array}\right)\in\Gamma\subset SL_2({\bf Z})$$
Then ${\cal F}(a,t_r)$ is defined as the formal solution to
\be
\frac{\partial{\cal F}}{\partial t_r}=-H^r\left(a,\frac{\partial{\cal F}}
{\partial a}\right)
\label{47}
\ee
where the Hamiltonians
$H^r$ are sort of specified below.  We do not identify the $t_r$ with
Whitham times here but observe that formally $d{\cal F}=
\sum a_i^Dda_i-\partial_n{\cal F}dt$ with $\partial{\cal F}/\partial a_i
=a_i^D$ and $-H_n=\partial_n{\cal F}$ establishes a connection.
One looks now at the geometric representation of data entering the 
construction of the low energy effective Abelian theory for general $r$. 
This is based
on \cite{sd} but reformulated in \cite{lg} in 
a manner related to our interests here.  Thus
let $\omega=\sum da^i\wedge da_i^D$ and $\theta=\sum_1^r a_i^Dda^i\equiv
(a^D,da)$.  Let $\Gamma$ be a subgroup of $Sp(2r,{\bf Z})$ and ${\cal L}$
a $\Gamma$ invariant Lagrangian submanifold in ${\bf C^{2r}}$.  By
definitions the restriction of $\omega$ to ${\cal L}$ vanishes so
$\theta|_{{\cal L}}=d{\cal F}$ where ${\cal F}:\,\,{\cal L}\to {\bf C}$.
This ${\cal F}$ is called a generating function of ${\cal L}$ and it is
globally well defined on ${\cal L}$ if ${\cal L}$ is simply connected.
${\cal F}$ transforms under the action of $g\in\Gamma$ via
\be
g=\left(\begin{array}{cc}
A & B\\
C & D
\end{array}\right);\,\,g^*{\cal F}(x)={\cal F}(x)+(Ba,Ca^D)+\frac{1}{2}(Ba,Da)
+\frac{1}{2}(Aa^D,Ca^D)+c(g)
\label{51}
\ee
where $c(g)$ is a certain cocycle $[c(g)]\in H^1(\Gamma,{\bf C})$.  If
$c(g)$ is trivial one can solve (\ref{51}) as
$
{\cal F}=(1/2)(a,a^D)+(u/\pi i)$
where $u$ is $\gamma$ invariant on ${\cal L}$ (cf. here \cite{ba,ea,ma,se}).  
This property reflects the scaling properties of ${\cal F}$.  To see this
insert $a^D=\partial{\cal F}/\partial a$ into this ${\cal F}$ and use (\ref{47})
(assuming the extension of $u$ to ${\bf C^{2r}}$ is known).  Then one
claims in \cite{lg} that the $\Gamma$ invariant ${\cal L}$ determines
an effective abelian $N=2$ gauge theory with duality group $\Gamma$.  Note
that ${\cal F}$ always means prepotential in \cite{lg} and generating function
is only used in the precise sense just indicated.
\\[3mm]\indent
Thus symplectomorphisms of ${\bf C^{2r}}$ map ${\cal L}$ to another
Lagrangian submanifold and the symplectomorphisms in the component of
the identity are generated by time dependent Hamiltonians $H(a,a^D,t)$ 
with local description
\be
\frac{\partial {\cal F}}{\partial t}=-H\left(a,\frac{\partial
{\cal F}}{\partial a},t\right)
\label{53}
\ee
It is worth comparing this to (\ref{RY}), (\ref{VY}), (\ref{jy}) etc.
for $w\sim {\cal F}$ to show
that one is essentially writing RG equations here 
with the $a_i$ as coupling constants.  The argument
in \cite{lg} is as follows. 
The flows which 
preserve $\Gamma$ invariance are generated by $\Gamma$ invariant $H$ and we 
let ${\cal C}$ denote the space of all $\Gamma$ invariant holomorphic functions
on ${\bf C^{2r}}$.  The Hamiltonian flows which do not change ${\cal L}$ 
are generated by Hamiltonians satisfying
\be
\tau_{ij}\frac{\partial H}{\partial a_i^D}=-\frac{\partial H}{\partial a_j};
\,\,\tau_{ij}=\frac{\partial^2{\cal F}}{\partial a_i\partial a_j}=
\frac{\partial a^D_j}{\partial a_i}
\label{54}
\ee
and the space of such Hamiltonians is called ${\cal C}_{{\cal L}}$.  Then
${\cal W}_{{\cal L}}={\cal C}/
{\cal C}_{{\cal L}}\sim\Gamma$ invariant functions on
${\cal L}$ and there is no canonical way of extending such functions to
$\Gamma$ invariant functions on ${\bf C^{2r}}$.  There are two possible
difficulties:  {\bf(A)} The Hamiltonians may be time dependent and
{\bf (B)} Even if time independent there are many ways to extend 
$u\in{\cal C}$ to ${\bf C^{2r}}$.  To dispose of these problems note that  
functions $H_k(a,a^D,t)$ can be used in defining a consistent system (\ref{53})
if and only if
$
(\partial H_k/\partial t_m)-(\partial H_m/\partial k_k)+
\{H_m,H_k\}=0$.
Therefore impose the extra condition $\partial_nH_k=0$ (background
independence) to take care of {\bf (A)}.  For {\bf (B)} take first 
$r=1$ with ${\bf C^*}$ acting in ${\bf C^2}$ in a standard way.  This
action commutes with $\Gamma$ and let $u$ be a basis of $\Gamma$ invariant
functions on ${\cal L}\,\,({\cal L}$ is one dimensional) with other
admissable functions being rational functions of $u$.  Then assume
homogeneity, namely $u$ extends to a $\Gamma$ invariant function
$H(a,a^D)$ on ${\bf C^2}$ with properties
\be
H(\mu a,\mu a^D)=\mu^dH(a,a^D);\,\,H(a,a^D)|_{{\cal L}}=u
\label{56}
\ee
Since the $H_k$ must Poisson commute they will be functions of the
Hamiltonian corresponding to $u$; hence in particular for a polynomial
$P,\,\,P(u)\to P(H(a,a^D))$.
The deformation problem is now well posed 
as in \cite{Az}; one has to solve
$
\dot{a}=(\partial H/\partial a^D)$ and $\dot{a}^D=-(\partial H/
\partial a)$.
with the initial conditions $(a(0),a^D(0))\in {\cal L}$.  Then
\be
{\cal F}(a,t)={\cal F}(\tilde{a},0)+\int^t_0[a^D(t')\dot{a}(t')-
H(a(t'),a^D(t'))]dt'
\label{58}
\ee
where the trajectory $(a(t'),a^D(t'))$ is such that $a(0)=\tilde{a}$
and $a(t)=a$ (note that this comes from $d{\cal F}=a^Dda-Hdt=(a^D\dot{a}
-H)dt$).  
Then one introduces the set of times
\be
\frac{\partial {\cal F}}{\partial t^k}=-H^k\left(a,\frac{\partial{\cal
F}}{\partial a}\right)
\label{59}
\ee
and this allows one to compute the ${\cal F}_{k_1,\cdots,k_p}$ in 
$${\cal F}
(a,t)={\cal F}_0(a)+\sum_{k>0}t^k{\cal F}_k+\sum_{k,\ell >0}\frac{1}{2}
t^kt^{\ell}{\cal F}_{k,\ell}+\cdots$$  In particular ${\cal F}_k$ depends only
on the restriction of $H^k$ to ${\cal L}$ while ${\cal F}_{k\ell}=$
pair contact term depends on the 1-jets of $H^k$ and $H^{\ell}$ via
\be
{\cal F}_{k\ell}=k\ell u^{k+\ell -2}\frac{\partial H}{\partial a^D}\frac
{du}{da}
\label{60}
\ee
where $du/da\sim$ derivative along ${\cal L}$.  By quasihomogeneity
of $H$ this yields
\be
{\cal F}_{k\ell}=k\ell u^{k+\ell -2}\left(\frac{du-a(du/da)}{a^D(da/du)
-a(da^D/du)}\right)
\label{61}
\ee
For $r>1$ there is a question of how to extend $u_1,\cdots, u_r$ where
for $G=SU(r+1)$ one has the SW curve and differential
\be
z+\frac{1}{4z}=x^{r+1}+\sum_1^ru_kx^{r+1-k};\,\,d\lambda=\frac{xdz}{z}
\label{CSW}
\ee
The paper \cite{lg} goes on to discuss many sophisticated matters 
but for our purposes it is enough to exhibit
HJ equations for ${\cal F}$ arising as 
in Section 2 for the RG with $w\sim {\cal F}+\Gamma\kappa^D$.
The connection here is somewhat nebulous at this point however.

\section{CONNECTIONS}
\renewcommand{\theequation}{8.\arabic{equation}}\setcounter{equation}{0}

Many discussions of SW theory simply ignore the $T_n$ Whitham variables
and concentrate on the $a_i,\,\,u_k,\,\,h_k,\,\,\Lambda$ etc.  This is
fine but nevertheless the Whitham dynamics is there for the asking
once the curves arise from an integrable system such as KP/Toda,
Calogero-Moser (CM), etc.  We have seen that the Whitham times $T_n$
(for some finite set) simply form a flat coordinate system on a moduli
space related to a TFT.  They form part of a larger set of moduli
$(a_i,T_n,\cdots)\sim (T_A)$ which describe the SW curve and gauge theory.
They correspond in some sense to adiabatic deformations which means that all
curve parameters (branch points, periods, differentials, etc.) depend on
the $T_n$ (along with moduli such as Casimirs $h_k$, etc.).  Actually
(cf. \cite{na}) the $a_i$ variables are associated to holomorphic differentials
$d\omega_i$ and can be inserted into BA functions as in \cite{cc,cg,na} to 
play a parallel role to the $T_n\sim d\Omega_n$.  However the $a_i$ and $a_i^D$ 
give rise to the physical spectrum of the theory while the role of the
$T_n$ is not clear (beyond deformation of curves
or coupling to gravity).  The temptation to
interpret the $d\Omega_n$ as chiral primary fields of one or two
$A_r$ type topological strings is mentioned in \cite{na} and this seems
to portray a topological deformation family of SW theories associated to a
given situation at say $T_n=0$.
Thus never
mind then the role of $T_n$ as moduli but think of them as deformation
parameters. They may not be RG parameters but they generate
similar flows.  In this spirit we think of $a_i(T),\,\,h_k(T),$ etc. and
with hindsight to Section 7 one can anticipate some sort of HJ equation
as in Section 2 (or 7) governing the flow of moduli under Whitham times.
In this context we have
\\[3mm]\indent {\bf THEOREM 8.1.}$\,\,$
The Whitham dynamics itself can furnish beta functions
for the deformation theory as in (\ref{11}) - (\ref{113}) 
or (\ref{119}) below. 
\\[3mm]\indent 
Indeed, in the context of \cite{ia} for example one
expressses the Whitham dynamics for the Casimirs $h_k$ via (assume here
the number $K$ of moduli $h_k$ equals the genus $g$)
\be
\sum\partial_nh_k\left(\sum T_m\sigma_{ki}^m\right)\equiv \sum\partial_n h_k
\sigma_{ki}=-c^n_i
\label{9}
\ee
where $d\hat{\Omega}_n=d\Omega_n+\sum_1^gc^n_id\omega_i$ and $(\partial
d\hat{\Omega}_m/\partial h_k)=\sum\sigma_{ki}^m(h)d\omega_i$.  Here the
difference between $d\hat{\Omega}_n$ and $d\Omega_n$ is simply that
$\oint_{A_i}d\Omega_n=0$.  This implies
\be
\frac{\partial h_k}{\partial T_n}=-\sum_1^g c^n_i(h)\sigma_{ik}^{-1}(h)
\label{11}
\ee
Here we recall
\be
\frac{\partial}{\partial h_k}\left(2{\cal F}-\sum a_j\frac{\partial{\cal F}}
{\partial a_j}\right)=\oint_{sing}{\cal S}
\frac{\partial d{\cal S}}{\partial h_k}
=
\label{8}
\ee
$$=\sum T_n\frac{\partial}{\partial h_k}\partial_n{\cal F}=
\sum T_n\partial_n\frac{\partial{\cal F}}{\partial h_k};\,\,\left(\sum a_j
\frac{\partial}{\partial a_j}+\sum T_n\frac{\partial}{\partial T_n}\right)
h_k=0$$
For $T_n=log(\kappa_n)$ one sees that (\ref{11}) provides a formula
for $\kappa_n(\partial h_k/\partial \kappa_n)$ which we call
$\beta_n^k$.  Note also that the generic Whitham equations $\partial_A
d\Omega_B=\partial_Bd\Omega_A$ would also provide a formula $\partial_n
d\Omega_1=\partial_1d\Omega_n$ for example along with
$\partial d\omega_i/\partial a_j=\partial d\omega_j/\partial a_i$ and
$\partial d\omega_i/\partial T_A=\partial d\Omega_A/\partial a_i$
(cf. \cite{cc,na}).
Now the format of Section 2, e.g. (\ref{VY}), gives (setting $t\sim T_n$
and $w\sim F+\Gamma\kappa_n^D$ with $F=F(h_k,\Lambda)$ and $D=K$)
\be
\partial_nF+\sum_1^K\beta_n^k\frac{\partial F}{\partial h_k}+\kappa_n^D
U_n(h_k)=0
\label{113}
\ee
Such an equation would define $U_n$ at least but we don't see an immediate
application. 
\\[3mm]\indent 
Take next $D=1$ with $F=F(a,\Lambda)$ and write
$\partial_nF+(\partial_na)F_a+\kappa_nU_n(a)=0$ where $a=a(T,\Lambda)$
(see below for dependencies).  This should hold along with
$\Lambda F_{\Lambda}+(\Lambda a_{\Lambda})F_a+\Lambda U_1(a)=0$.
For genus one we could also take $\tau$ as the modulus to obtain
$(\bullet\clubsuit)\,\,
\Lambda F_{\Lambda}+(\Lambda\tau_{\Lambda})F_{\tau}+\Lambda\hat{U}_1(\tau)=0$.
In the situation $(\bullet
\clubsuit)$ one knows from \cite{ba} that for $G(\tau)=
(u/\Lambda^2)\,\,(\sim G_3(\tau)$) one has $\beta(\tau)=\Lambda\partial_
{\Lambda}\tau=-2G/G'$ so putting $\Lambda\partial_{\Lambda}F=2u/i\pi$
in $(\bullet\clubsuit)$ we obtain ($'\sim\partial_{\tau}$)
\be
\Lambda\hat{U}_1(\tau)=-\frac{2u}{i\pi}+\frac{2GF'}{G'}=
-\frac{2u}{i\pi}-\beta F'
\label{117}
\ee
which
could be written in terms of $\beta$ and $F'=F_{\tau}$ using
\be
u(\tau)=\left(\frac{\Lambda}{\Lambda_0}\right)^2u(\tau_0)
exp\left(-2\int_{\tau_0}^{\tau}\beta^{-1}(x)dx\right)
\label{118}
\ee
from \cite{ba}. This doesn't seem to lead to any new conclusions however.
Generally as in \cite{ba} one can consider $a=a(u,\Lambda),\,\,F=
F(a,\Lambda),\,\,\tau=F_{aa}=\tau(a,\Lambda),\,\,(u/\Lambda^2)=G_1(a)=G_3(\tau),
\,\,u=u(a,\Lambda),$ etc. (omitting any $T_n$ dependence).  After turning
on the Whitham dynamics (or deformation theory) one obtains $h_k=h_k(T_n),
\,\,u=u(T_n),\,\,a_k=a_k(T_n)$,
etc. and in \cite{cc,cg,ia} one shows how to develop
$a_i$ and $T_n$ as independent variables and sort out the $T_n$ dependencies.
In that situation $a^D_i$ will depend on the $T_n$ (cf. \cite{cc})
and (\ref{113}) would have an analogue  ($D=g$)
\be
\partial_nF+\sum_1^g\hat{\beta}_n^k\frac{\partial F}{\partial a^D_k}+
\kappa_n^D\hat{U}_n=0;\,\,\hat{\beta}_n^k=\kappa_n\frac{\partial a^D_k}
{\partial\kappa_n}
\label{119}
\ee
Thus we have shown that ``Hamiltonians" of the form
$\sum\beta^a\phi_a\sim \sum(\partial_tg^a)(\partial w/
\partial g^a)$, coupled to $\partial_tw$ in HJ type equations, arise
naturally in both RG theory and in the deformation of $N=2$ susy gauge
theories by Whitham dynamics.


\begin{thebibliography}{cc}


%
\bibitem{Aa} S. Aoyama and Y. Kodama,
hep-th 9505122, Comm. Math. Phys., 182 (1996), 185-219
%
\bibitem{Az} V. Arnold,
Mathematical methods in classical mechanics, Springer, 1980
%
\bibitem{bc} D. Bellisai, F. Fucito, M. Matone, and G. Travaglini,
hep-th 9706099
%
\bibitem{bh} E. Belokolos, A. Bobenko, V. Enolskij, A. Its, and V. Matveev,
Algebro-geometric approach to nonlinear integrable equations, Springer, 1994
%
\bibitem{By} G. Bertoldi and M. Matone,
hep-th 9712039 and 9712109
%
\bibitem{bl} J. Bie\'nkowska,
hep-th 9109003
%
\bibitem{bb} A. Bloch and Y. Kodama,
SIAM Jour. Appl. Math., 52 (1992), 909-928
%
\bibitem{ba} G. Bonelli and M. Matone,
Phys. Rev. Lett., 76 (1996), 4107-4110;
Phys. Rev. Lett., 77 (1996), 4712-4715
%
\bibitem{Bz} G. Bonelli and M. Matone,
hep-th 9712025
%
\bibitem{be} G. Bonelli, M. Matone, and M. Tonin,
Phys. Rev. D, 55 (1997), 6466-6470
%
\bibitem{bz} L. Brown, 
Ann. Phys. 126 (1980), 135-153; (with J. Collins) Ann. Phys., 130 (1980), 
215-248
%
\bibitem{cc} R. Carroll and J. Chang,
solv-int 9612010, Applicable Anal., 64 (1997), 343-378
%
\bibitem{cg} R. Carroll,
solv-int 9606005, Proc. Second World Congress Nonlin. Analysts,
Athens, 1996, Nonlin. Anal., 30 (1997), 187-198
%
\bibitem{ch} R. Carroll and Y. Kodama,
Jour. Phys. A, 28 (1995), 6373-6387
%
\bibitem{ci} R. Carroll,
Jour. Nonlin. Sci., 4 (1994), 519-544; Teor. Mat. Fizika, 99 (1994), 220-225
%
\bibitem{cj} R. Carroll,
Proc. NEEDS'94, World Scientific, 1995, pp. 24-33;
Repts. Math. Phys., 37 (1996), 1-21
%
\bibitem{cw} R. Carroll,
solv-int 9703013, Phys. Lett. A, 234 (1997), 171-180
%
\bibitem{cm} R. Carroll,
Applicable Anal., 49 (1993), 1-31; 56 (1995), 147-164
%
\bibitem{ca} B. Craps, F. Roose, W. Troost, and A. vanProeyen,
hep-th 9703082
%
\bibitem{dy} K. Davis,
hep-th 9308039
%
\bibitem{dz} E. D'Hoker, I. Krichever, and D. Phong,
hep-th 9610156, Nucl. Phys. B, 494 (1997), 89-104;
hep-th 9609041 and 9609145
%
\bibitem{dg} E. D'Hoker and D. Phong,
hep-th 9701055 and 9709053
%
\bibitem{Dz} L. Dickey,
Soliton equations and Hamiltonian structure, World Scientific, 1991
%
\bibitem{dw} R. Dijkgraaf, E. Verlinde, and H. Verlinde,
Nucl. Phys. B, 352 (1991), 59-86;
Nucl. Phys. B, 348 (1991), 435-456
%
\bibitem{di} B. Dolan,
Inter. Jour. Mod. Phys. A, 9 (1994), 1261-1286
%
\bibitem{dv} B. Dolan,
hep-th 9307023, 9307024, and 9403070; 
cond-mat 9412031
%
\bibitem{do} B. Dolan,
hep-th 9702156 and 9710161
%
\bibitem{De} B. Dolan,
hep-th 9406061
%
\bibitem{dp} B. Dolan,
hep-th 9511175, Inter. Jour. Mod. Phys. A, 12 (1997), 2413-2424
%
\bibitem{de} R. Donagi,
alg-geom 9705010
%
\bibitem{da} R. Donagi and E. Witten,
Nucl. Phys. B, 460 (1996), 299-334
%
\bibitem{dc} B. Dubrovin,
Lect. Notes Math., Springer, 1996, pp. 120-348;
Nucl. Phys. B, 379 (1992), 627-689
%
\bibitem{Dy} B. Dubrovin and S. Novikov,
Russ. Math. Surveys, 44:6 (1989), 35-124
%
\bibitem{ea} T. Eguchi and S. Yang,
Mod. Phys. Lett. A, 11 (1996), 131-138
%
\bibitem{eb} T. Eguchi, Y. Yamada, and S. Yang,
Mod. Phys. Lett. A, 8 (1993), 1627-1637
%
\bibitem{fc} H. Flaschka, M. Forest, and D. McLaughlin,
Comm. Pure Appl. Math., 33 (1980), 739-784
%
\bibitem{fg} P. Fr\'e and P. Soriani,
The $N=2$ wonderland, World Scientific, 1995
%
\bibitem{fb} F. Fucito, A. Gamba, M. Martellini, and O. Ragnisco,
Inter. Jour. Mod. Phys. B, 6 (1992), 2123-2147
%
\bibitem{ff} F. Fucito and G. Travaglini,
Phys. Rev. D, 55 (1997), 1099-1104
%
\bibitem{gh} J. Gaite and D. O'Connor,
hep-th 9511092
%
\bibitem{gc} A. Gorsky, I. Krichever, A. Marshakov, A. Mironov, and
A. Morozov,
Phys. Lett. B,355 (1995), 466-474
%
\bibitem{gd} A. Gorsky,
hep-th 9612238
%
\bibitem{ge} A. Gorsky, S. Gukov, and A. Mironov,
hep-th 9707120 and 9710239
%
\bibitem{gg} A. Gorsky and A. Marshakov,
Phys. Lett. B, 375 (1996), 127-134
%
\bibitem{Ga} B. Greene,
hep-th 9702155
%
\bibitem{gb} P. Guha and K. Takasaki,
solv-int 9705013
%
\bibitem{gk} A. Gurevich and L. Pitaevskij,
JETP, 65 (1973), 590-604
%
\bibitem{hh} J. Harnad,
hep-th 9406078; solv-int 9710012 and 9710016
%
\bibitem{he} P. Howe and P. West,
Nucl. Phys. B, 486 (1997), 425-442
%
\bibitem{iz} K. Ito and S. Yang,
hep-th 9712018, 9708017, and 9603073; Phys. Lett. B, 366 (1996), 165-173
%
\bibitem{ia} H. Itoyama and A. Morozov,
hep-th 9511126, 9512161, and 9601168; Nucl. Phys. B, 477 (1996), 855-877 and
491 (1997), 529-573
%
\bibitem{kr} A. Klemm,
hep-th 9705131
%
\bibitem{kd} A. Klemm, W. Lerche, and S. Theisen,
Inter. Jour. Mod. Phys. A, 11 (1996), 1929-1973
%
\bibitem{ke} Y. Kodama and J. Gibbons,
Workshop on nonlinear and turbulent processes in physics, World Scientific,
1990, pp. 166-180
%
\bibitem{ka} I. Krichever and D. Phong,
hep-th 9604191, Jour. Diff. Geom, 45 (1997), 349-389
%
\bibitem{kb} I. Krichever,
Funkts. Anal. Prilozh., 22 (1988), 200-213
%
\bibitem{kc} I. Krichever,
Comm. Pure Appl. Math., 47 (1994), 437-475;
Acta Applicandae Math., 39 (1995), 93-125
%
\bibitem{ko} I. Krichever and D. Phong,
hep-th 9708170
%
\bibitem{kz} I. Krichever,
hep-th 9611158
%
\bibitem{Ka} T. Kubota and N. Yokoi,
hep-th 9712054
%
\bibitem{le} M. L\"assig,
Nucl. Phys. B, 334 (1990), 652-668
%
\bibitem{lf} J. Latorre and C. L\"utken,
hep-th 9711150
%
\bibitem{la} W. Lerche,
hep-th 9611190
%
\bibitem{lg} A. Losev, N. Nekrasov, and S. Shatashvili,
hep-th 9711108
%
\bibitem{md} A. Marshakov,
hep-th 9602005, 9702083, and 9709001; Mod. Phys. Lett. A, 11 (1996),
1169-1183
%
\bibitem{mf} A. Marshakov, M. Martellini, and A. Morozov,
hep-th 9706050
%
\bibitem{My} A. Marshakov, A. Mironov, and A. Morozov,
hep-th 9607109 and 9701123; Mod. Phys. Lett. A, 12 (1997), 773-787
%
\bibitem{me} E. Martinec,
Phys. Lett. B, 367 (1996), 91-96
%
\bibitem{mg} E. Martinec and N. Warner,
Nucl. Phys. B, 459 (1996), 97-112;
hep-th 9511052
%
\bibitem{ma} M. Matone,
Phys. Lett. B, 357 (1995), 342-348;
Phys. Rev. D, 53 (1996), 7354-7358;
Phys. Rev. Lett., 78 (1997), 1412-1415
%
\bibitem{Mq} J. Minahan, D. Nemeschansky, and N. Warner,
hep-th 9710146
%
\bibitem{Mw} A. Mironov,
hep-th 9704205
%
\bibitem{Ms} A. Mironov and A. Morozov,
hep-th 9712177
%
\bibitem{Mz} A. Morozov,
hep-th 9711194
%
\bibitem{mu} R. Myers and V. Periwal,
hep-th 9611132
%
\bibitem{na} T. Nakatsu and K. Takasaki,
Mod. Phys. Lett. A, 11 (1996), 157-168
%
\bibitem{oc} D. O'Connor and C. Stephens,
hep-th 9304095, 9310086, and 9310198
%
\bibitem{rd} A. Ritz,
hep-th 9710112
%
\bibitem{sd} N. Seiberg and E. Witten,
Nucl. Phys. B, 426 (1994), 19-52
%
\bibitem{sk} N. Seiberg,
Phys. Lett. B, 206 (1988), 75-80; 318 (1993), 469-475
%
\bibitem{se} J. Sonnenschein, S. Theisen, and S. Yankielovicz,
Phys. Lett. B, 367 (1996), 145-
%
\bibitem{Sb} H. Sonoda,
Nucl. Phys. B, 352 (1991), 585-600 and 601-615; hep-th 9306119
%
\bibitem{sa} C. Stephens,
hep-th 9611062
%
\bibitem{ta} K. Takasaki and T. Takebe,
Inter. Jour. Mod. Phys. A, Supp. 1992, pp. 889-922; Rev. Math. Phys.,
7 (1995), 743-808
%
\bibitem{tc} K. Takasaki and T. Nakatsu,
hep-th 9603069
%
\bibitem{td} K. Takasaki,
solv-int 9704004 and 9705016; hep-th 9711058
%
\bibitem{wa} E. Witten,
Jour. Geom. Phys., 8 (1992), 327-334
%
\bibitem{ya} T. Yoneya,
Comm. Math. Phys., 144 (1992), 623-639
%
\bibitem{zc} M. Zabzine,
hep-th 9705015 and 9707064
%
\bibitem{zb} A. Zamolodchikov,
JETP Lett. 43 (1986), 730-732; Sov. Jour. Nucl. Phys., 44 (1986), 529-533;
46 (1987), 1090-1096
%

\end{thebibliography}
\end{document}